# Green Ammonia:
# A Techno-Economic Supply Chain Optimization

Lucien Genge[a,*], Felix Müsgens[a]

[a] *Brandenburg University of Technology, Siemens-Halske-Ring 13, Cottbus, Germany*

* *lucien.genge@b-tu.de*, *(corresponding author)*

# ABSTRACT

Green ammonia is emerging as a strategic intermediary within green energy supply chains, serving effectively as both an industrial commodity and hydrogen carrier. This study provides a techno-economic analysis of green ammonia supply chains, comparing cost-effective pathways from global production to European consumers, and evaluates ammonia alongside alternative hydrogen carriers. Gaseous hydrogen consistently remains the most economical import option for Europe, though ammonia holds a narrowing cost advantage over liquid hydrogen (from 16 % in 2030 to 10 % by 2040). Competitive ammonia suppliers, notably Morocco, the United States, and the United Arab Emirates, benefit from low renewable energy costs, with significant price reductions expected by 2040, driven by falling costs for electricity, electrolysers, and conversion technologies. Optimal transport modes vary by consumer demand and distance: trucks are ideal for low demands at all distances, rail for medium ranges, and pipelines for high-demand scenarios. By 2040, ammonia will primarily serve direct-use applications, as hydrogen consumers increasingly shift to direct hydrogen supplies. Policymakers should prioritize pipeline infrastructure for hydrogen distribution, cautiously invest in ammonia's short- to medium-term infrastructure advantages, and limit long-term reliance on ammonia as a hydrogen carrier to mitigate stranded asset risks.

# 1 INTRODUCTION

The transition to green energy is critical to achieve the European Union's climate targets and diversify its energy supply. Estimates indicate that approximately 90 % of global greenhouse gas emissions are due to fossil fuel use [1]. To significantly reduce GHG emissions, the current fossil-based energy and industrial systems must be decarbonized. The shift towards renewable energy sources, also known as renewables, is essential and increasingly cost-effective [2], [3], [4], [5]. While the power sector is transitioning relatively well with the rapid adoption of photovoltaics and wind energy, sectors such as long-range transportation and industry may need to rely commodities which are based on green hydrogen [6], [7].

Ammonia already is an important commodity in agriculture and basic chemical industries. Known for its extensive global trading history and established regulatory frameworks, green ammonia not only serves as an alternative to fossil-based ammonia but also plays a crucial role in the broader hydrogen economy [8]. Under the EU's RePowerEU plan [9], green ammonia is a key candidate to meet the ambitious target of consuming 660 TWh of green commodities by 2030, with half expected to be sourced through imports. In Germany, where demand for hydrogen is projected to far exceed domestic production capacity [10], green ammonia imports offer a promising solution. Currently, global ammonia production, which is primarily for fertilizer production, stands at 971 TWh annually with 10 % being cross-regionally transported [11]. This is expected to surge to 1,575 TWh by 2050 as its applications expand, including its potential use as a hydrogen carrier [12].

Despite the importance of ammonia, current research lacks a comprehensive analytical approach that covers supply costs from international production sites to final consumer locations. Especially ammonia's dual role as both a direct commodity and a transport medium for hydrogen is often neglected. Against this background our paper is the first to introduce an approach that calculates costs from international production sites to domestic consumption sites. Specifically, we parameterize supply chain models for ten export countries to estimate Well-to-Border import prices and evaluate Border-to-Consumer distribution costs, highlighting ammonia's adaptability in the medium-term energy transition. Further, we provide quantitative insights into these import prices for ammonia, liquid hydrogen, and gaseous hydrogen, and determine cost-optimal inland distribution pathways, initially applied to 100 generic European consumer sites and subsequently demonstrated through 14 empirical ammonia and hydrogen consumer sites in Germany. Additionally, our research identifies the most cost-effective global suppliers, project substantial cost reductions by 2040, and reveal strategic advantages for infrastructure and distribution mode investments, supporting ammonia's dual role as a direct commodity and hydrogen carrier.

Following this introduction, our paper includes a detailed Literature Review that explores various studies on green commodity imports and inland distribution as well as identifies the research gaps. This sets the stage for the Modelling the Ammonia Supply Chain chapter, which outlines the two models, the assumptions made, and the data utilized. Results & Discussion chapter presents an in-depth analysis of import prices at the border, and supply costs at consumer site. Finally, the Conclusions are drawn in the last chapter.

## 2 LITERATURE REVIEW

While numerous studies have examined ammonia supply chains, significant research gaps remain that hinder a comprehensive understanding of its role in the green energy transition. These gaps arise from the fragmented approach in existing literature, which either focuses on import cost estimation or inland distribution but rarely integrates both perspectives. Furthermore, most studies calculate supply costs at the border but fail to consider the more accurate prices that consumers will face. This section identifies three major areas requiring further investigation.

### 2.1 Import Costs

Ammonia import costs are shaped by a combination of production costs and transport cost with significant regional variations influencing its competitiveness as a globally traded energy commodity.

Several studies have assessed production costs in regions with abundant renewables, highlighting differences in cost competitiveness. Fasihi et al. [13] identify Patagonia, the Atacama Desert, Tibet, and the Horn of Africa as the most cost-effective production regions, with costs falling below 113 €/MWh in 2020 and projected to decline to 49 €/MWh by 2050 due to technological advancements. However, Nayak-Luke and Bañares-Alcántara [14] argue that green ammonia is not yet competitive with fossil-based ammonia, though cost parity is expected in the 2030s, particularly in Northern Africa, where government support may be required to lower costs. Additionally, Armijo and Philibert [15] demonstrate that in Chile and Argentina, hybrid renewable systems (solar and wind) could reduce ammonia production costs by 5–13 %. Despite these cost reductions, the geographic mismatch between production and consumption centres introduces additional transportation costs, which are critical in determining ammonia's competitiveness both as feedstock and as a hydrogen carrier.

Further studies have examined the role of transport costs in shaping ammonia's economic feasibility for international trade. Egerer et al. [16] model ammonia imports from Australia to Germany, concluding that by 2030, green ammonia could be cost-competitive with grey ammonia, particularly if CO2 pricing is introduced, though transport and storage costs remain key factors. Ishimoto et al. [17] emphasize that while reconverting ammonia to hydrogen incurs additional costs, ammonia's lower transport costs and existing infrastructure make it a cost-effective direct energy carrier. This is supported by Hampp et al. [18] and Agyekum et al. [19], who find that imported ammonia is cheaper than domestic production, with shipping costs lower than those for liquid hydrogen. Hank et al. [20] further confirm ammonia's transport cost advantage, identifying it as one of the lowest-cost energy carriers for routes such as Morocco to Germany. Across studies, South America, the Middle East, North Africa, and Australia consistently emerge as the most competitive exporting regions. However, many studies examine export regions in isolation (e.g., [16], [20], [21], [22], [23], [24], [25], [26], [27], [28], [29]) rather than within a global cost-competitive framework, limiting the ability to determine marginal suppliers and prices.

While existing studies provide valuable insights into ammonia production and import costs, they often report levelized costs rather than actual market prices. Moritz et al. [21] and Galimova et al. [29] conclude that Germany could import green ammonia more cost-efficiently than producing it domestically, but their analysis clusters supplier countries instead of constructing a global supply curve. Without a merit-order approach ranking suppliers by cost, studies fail to identify marginal suppliers which are those with the highest costs that determine market prices. Salmon et al. [30] optimize a global ammonia network but do not explicitly model price-setting dynamics, instead predicting that green ammonia could be the cheapest option by 2050, although with regional cost disparities. Galimova et al. [31] identify ammonia as a key energy commodity, estimating that 29 % of global ammonia demand by 2050 will be met through international trade. While these studies use a cost-based merit-order approach, they still report levelized costs rather than price estimates. Genge et al. [32] underscore this gap through a meta-analysis of 30 import cost studies, revealing that ammonia import costs at the European border in 2030 range from 50 to 204 €/MWh. However, a more reliable price reference comes from the latest H2Global green ammonia auction, in which an Egyptian production site secured a contract to supply 2 TWh by 2033 at a price of 192 EUR/MWh [33]. This price is at the upper end of the meta-analysis spectrum, suggesting that market prices tend to be higher than cost-based estimates, emphasizing the need for more realistic price modelling in future research.

## 2.2 Inland Distribution Costs

Although ammonia import costs have been widely studied, inland distribution remains an overlooked factor, despite its significant impact on total supply costs. Many consumers are not located at import terminals and require additional transport via pipeline, rail, or truck. While long term plans such as the European hydrogen backbone aim to facilitate hydrogen transport, ammonia's role in inland distribution, especially in the medium term, remains uncertain.

Extensive research exists on inland hydrogen distribution, covering various transport modes including trucks, rail, and pipelines [34], [35], [36], optimal hydrogen carriers, comparing gaseous hydrogen, liquid hydrogen, and liquid organic hydrogen carriers [34], [36] , and distribution costs dependent on distance and demand [37], [38]. However, a similarly detailed analysis for ammonia is notably absent from the literature. Nayak-Luke and Banares-Alcantara [14] provide a foundational framework by outlining the key components for a techno-economic analysis of ammonia transport via truck, rail, and ship. While they reported relevant cost parameters, their study lacks a detailed cost analysis of those.

Other studies have focused on specific applications of ammonia for decarbonizing particular sectors, such as short-distance shipping [39] and railway freight transport [40], but do not provide a comprehensive cost comparison across transport modes. Cui and Aziz [41] analysed levelized costs of hydrogen transport via ammonia and methanol, comparing truck, rail, pipeline, and ship as a function of distance and demand. Notably, they were the first to model inland transport units as discrete integers, improving accuracy. However, their study focused only on hydrogen consumers, neglecting direct ammonia applications and overall supply chain costs. Building on Cui and Aziz [41], Sanchez et al. [42] computed the total cost of supplying fuel cells using four transport modes (truck, rail, pipeline, and ship) for hydrogen, ammonia, and methanol. Their analysis covered varying transport distances up to 3,000 km, but only for a fixed demand level. By linearizing transport costs, they simplified Cui and Aziz's approach, making it more applicable for supply chain modelling, yet limited to single-consumer efficiency studies rather than broader distribution networks.

## 2.3 Integrated Supply Chain Analysis

The literature predominantly analyses either ammonia's international supply costs (Well-to-Border) or its inland distribution costs (Border-to-Consumer) but rarely integrate both perspectives into a comprehensive supply chain analysis. Several hydrogen-focused studies attempt comprehensive well-to-consumer analyses, incorporating production and transport elements. Collis and Schomäcker [43] using a Monte Carlo simulation to determine the delivered cost of renewable hydrogen globally, identifying the most cost-effective production locations and transport routes from nearly 6,000 sites. Case studies for Cologne, Germany, and Houston, U.S. show that northern Egypt and southern France (pipelines) provide the cheapest hydrogen for Cologne (228–282 €/MWh), while the western Gulf of Mexico (trucking) and southern California (pipelines) offer the lowest costs for Houston (228-258 €/MWh), with production costs expected to decline by 2050. Kim et al. [44] analysing the economic feasibility and carbon footprint of importing hydrogen via liquid hydrogen, liquid organic hydrogen carriers, and ammonia, assessing transport and distribution options. A case study on hydrogen imports from Indonesia to Korea shows that all transport modes remain below Korea's 2030 cost target, with pipelines being the most cost-effective for large-scale distribution, while trailers are viable for deliveries under 150 GWh. Sens et al [27] evaluating the energy efficiency and cost of green hydrogen supply chains from well to tank for fuel cell heavy-duty vehicles in Germany in 2030 and 2050, comparing compressed gaseous hydrogen, liquid hydrogen, liquid organic hydrogen carriers, methanol, and ammonia from domestic and international sources (North/Central Germany, Tunisia, and Argentina). The results show that supply chains maintaining hydrogen in its final state of use are the most cost-effective, with compressed gaseous hydrogen at 150 €/MWh (2030) and 120 €/MWh (2050), and liquid hydrogen at 210 €/MWh (2030) and 180 €/MWh (2050), while LOHC incurs high costs, particularly over long distances. Similarly, Restelli [45] evaluates the costs of delivering green hydrogen via ammonia and liquid hydrogen from North Africa to Northern Italy for hydrogen consumers in the industry and mobility sector. The study highlights that centralized cracking of ammonia at the port minimizes distribution costs, particularly for industrial sectors, by reducing the need for further inland transport of gaseous hydrogen. In contrast, decentralized cracking at refuelling stations increases transportation costs due to the higher number of trucks required to distribute gaseous hydrogen.

Although these studies combine international imports with inland distribution, they present two key limitations: First, they fail to consider ammonia's role for direct consumers, focusing solely on hydrogen applications. Second, they simplify inland transport modelling, using fixed demand and distances rather than accounting for variations in consumer locations and infrastructure constraints.

## 2.4 Research Questions

Based on the identified research gaps, this study addresses the following three key research questions:

(i) What are the marginal import prices for ammonia, liquid hydrogen, and gaseous hydrogen at the European border, and how are these expected to develop by 2040?

(ii) What are the optimal inland transport modes for ammonia and hydrogen consumers, considering varying distances and demand profiles?

(iii) What is ammonia's role in medium-term energy supply chains, both as a direct commodity and as a hydrogen carrier, and what strategic infrastructure implications arise from this dual application?

# 3 MODELLING THE AMMONIA SUPPLY CHAIN

This chapter outlines the methodology employed in this study, starting with the introduction of the modelling framework and a detailed description of two optimization models.

## 3.1 Modelling Framework

We divide the well-to-consumer supply chain into two distinct models with prices at the national border as the link. First, the Well-to-Border Model covers the production and international transport of green ammonia and hydrogen to the border of the importing country. While existing studies have modelled this phase, we include it to ensure consistency, enable comparison, and integrate specific assumptions. Second, the Border-to-Consumer Model focuses on inland distribution, considering consumer site parameters such as demand levels and transportation distances. Both models, along with all underlying data and assumptions, are open-source and available on GitHub[1].

---

[1] https://github.com/LuGenge/well-to-consumer-ammonia-model

Taken together, these two models determine the costs of various hydrogen-based energy carriers at the consumption site. We distinguish between consumptions sites for ammonia and hydrogen. For each of these final products, six different supply chains emerge, varying by imported commodity and transport mode.

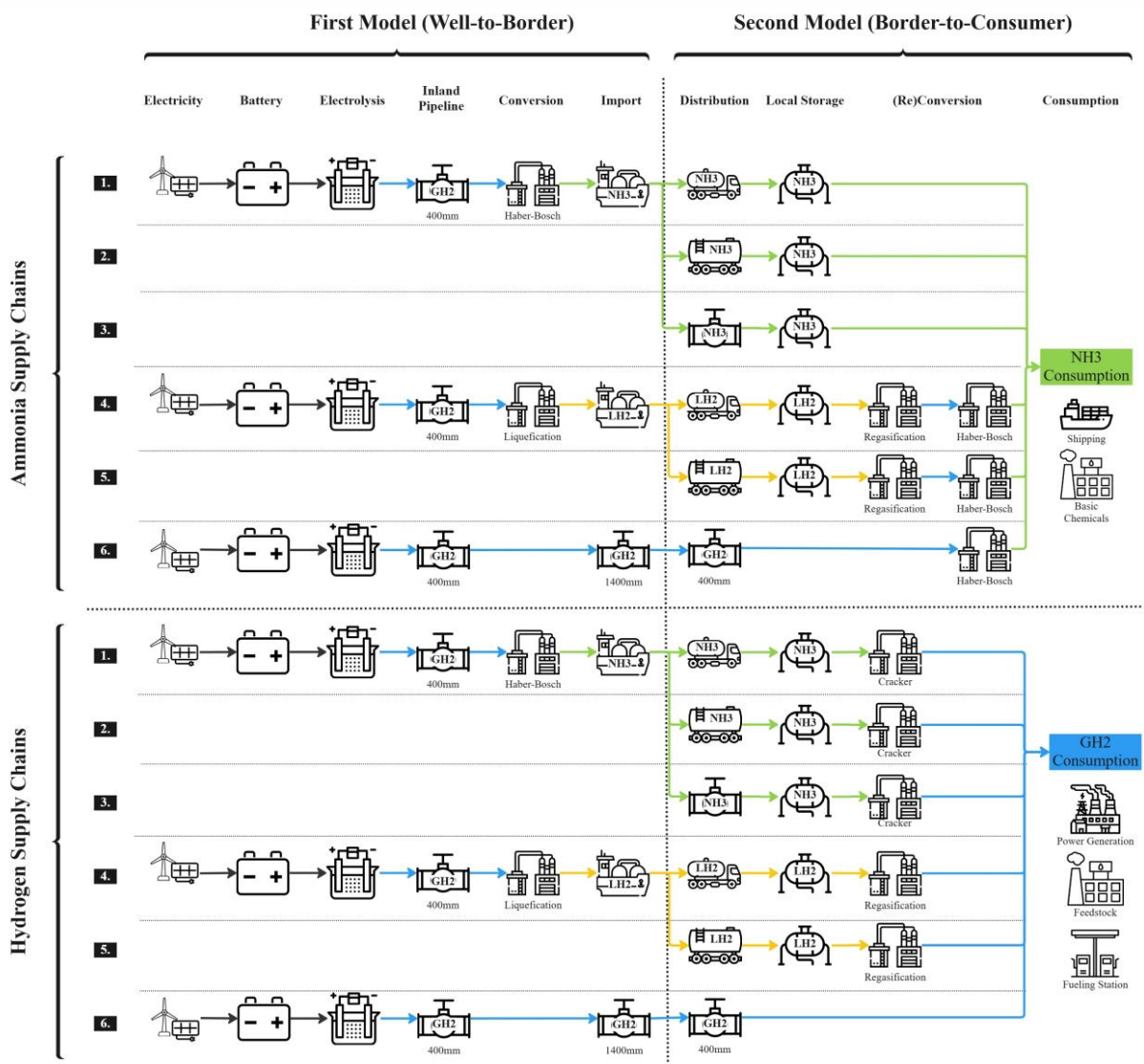

**Figure 1: Schematic visualization of assessed supply chains for ammonia and hydrogen consumers. Coloured arrows represent the specific commodity being transferred between processes: green for ammonia, yellow for liquid hydrogen, and blue for gaseous hydrogen.**

Figure 1 illustrates the supply chain structure for each type of consumptions site. The Well-to-Border model includes renewable electricity generation, hydrogen production via electrolysis, and conversion to liquid hydrogen or ammonia. Transport occurs via smaller inland pipelines, followed by international transport by ship or large pipelines. The model is formulated as a cost minimization problem, the resulting commodity prices at the border of the importing country are the key result for our analysis. The model will be described in detail in section 3.2. The second model uses the commodity prices as input and optimizes inland transport, storage, and conversion processes to match consumer requirements. Transport modes include pipelines, trucks, and rail, with final conversion steps dictated by the required product. The result of this second optimization model is the site-specific minimal cost of providing the required energy carrier. We included two consumer types: Ammonia consumers utilize ammonia directly. Hydrogen must first be converted via the Haber-Bosch process, and liquid hydrogen must be regasified before conversion. Ammonia consumers include inland shipping, basic chemical industries, and the fertilizer industry. Hydrogen consumers can use hydrogen directly from pipelines or rely on liquid hydrogen or ammonia, which require reconversion through regasification or cracking. These include power generation, steelmaking, chemical and refinery processes, and fuelling stations for heavy-duty vehicles. The model will be described in detail in section 3.3.

## 3.2 Well-to-Border Model

This section explains the modelling approach for determining import prices of commodities from well to border. It starts with a model overview, followed by the nomenclature and finally describes the model algebra.

### 3.2.1 Well-to-Border Model Overview

The Well-to-Border Model determines the import prices of commodities from production sites to the import border.

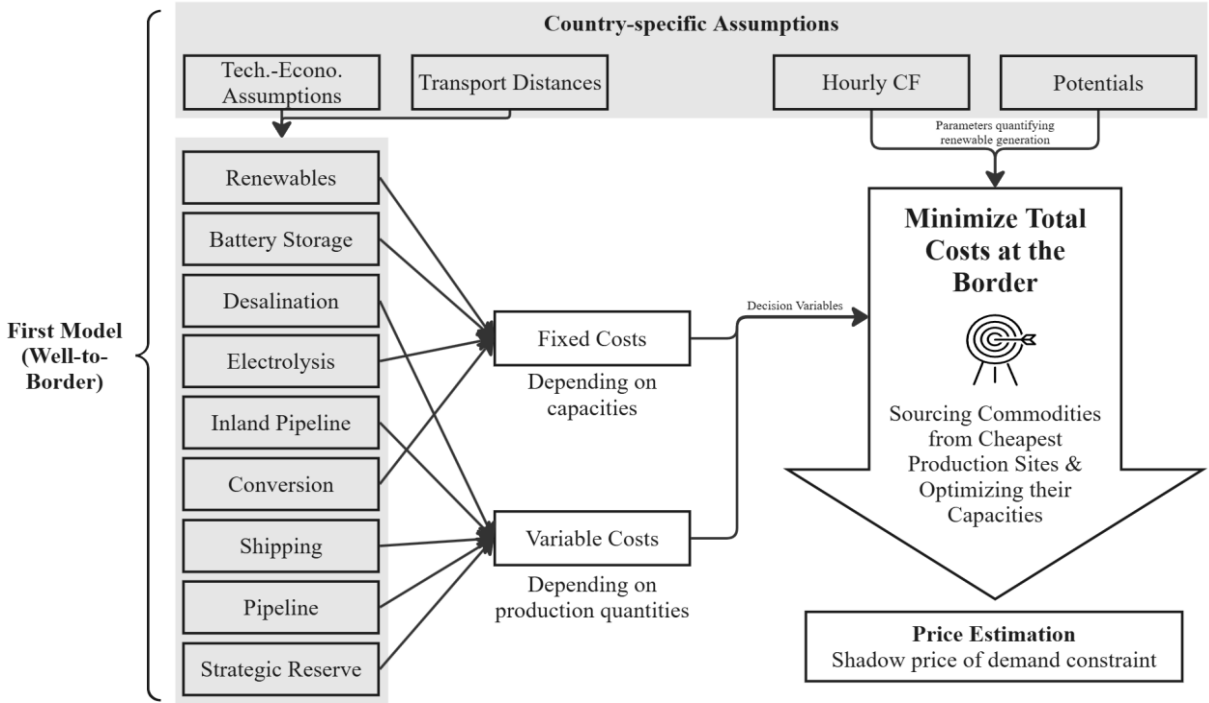

**Figure 2 Methodology for calculation the import price of commodities at the border.**

Figure 2 illustrates the methodology used for determining these import prices, focusing on cost categorization, optimization, and price estimation.

Cost categorization distinguishes between parameters and variables. Parameters, highlighted in grey, include specific investment and operational costs as model inputs. The following variables fixed costs and variable costs are optimized in the model. Fixed costs are computed from the combination of endogenously determined installed capacities (variables) and predefined specific investment costs for renewable energy generation, battery storage, electrolysis, and conversion processes (parameters). Variable costs depend on the operational parameters associated with production outputs, including expenses related to inland and international transportation, and are multiplied by the corresponding production quantities (variables).

The optimization model identifies cost-effective production sites by optimizing capacities and sourcing commodities from the cheapest locations. It ensures that fixed and variable costs are minimized while meeting demand constraints.

The commodity price at the border is derived from the shadow price of the demand constraint. In simple terms, this answers the question how much it would cost to provide one additional unit of supply.

### 3.2.2 Nomenclature

Note that we use capital letters for variables and small letters for sets and parameters.

| *Abbreviation* | *Unit* | *Description* |
|---|---|---|
| *Sets* | | |
| $n$ | | *Countries* |
| $r$ | | *Renewables* |
| $c$ | | *Resource classes* |
| $d$ | | *Distance to shore* |
| $i$ | | *Commodity* |
| $y$ | | *Years* |
| $h$ | | *Hours* |
| *Parameters* | | |
| $c_{n,r,y}^{RES}$ | *€/MW* | *Country-specific yearly costs of renewables.* |
| $c_{n,y}^{Batt}$ | *€/MW* | *Country-specific yearly costs of battery storage.* |
| $c_{n,y}^{El}$ | *€/MW* | *Country-specific yearly costs of electrolysis.* |
| $c_{i,y}^{Conv}$ | *€/MW* | *Country-specific yearly costs of commodities conversion.* |
| $c^{Desal}$ | *€/MWh* | *Costs of desalination and water transport.* |
| $c_{n,d,y}^{Tr,inl}$ | *€/MWh* | *Country-specific yearly costs of inland transport within the exporting country.* |
| $c_{n,i,y}^{Tr,int}$ | *€/MWh* | *Costs of international transport.* |

| $\eta_y^{El}$ | MW$_{H2}$/MW$_{el}$ | Efficiency of converting electricity to hydrogen. |
|---|---|---|
| $\eta_i^{Conv}$ | MW$_{Conv}$/MW$_{H2}$ | Efficiency of converting hydrogen to commodity. |
| $q_i^{Conv}$ | MW$_{el}$/MW$_{Conv}$ | Power demand of the commodities conversion process. |
| $\eta_{i,y}^{Conv,new}$ | MW$_{H2}$/MW$_{el}$ | Combined efficiency of hydrogen conversion and power demand. |
| $\eta^{Batt}$ | MW$_{el}$/MW$_{el}$ | Battery efficiency. |
| $pot_{n,r,c,d}^{RES}$ | MW | Potential capacity of renewables r in country n, resource class c, and distance to shore d. |
| $cf_{n,r,c,d}^{RES}$ | | Capacity factors of renewables r depending on the country n, resource class c, and distance to shore d. |
| $dem_y$ | MWh | European demand in year y. |

| ***Variables*** | | |
|---|---|---|
| $TC_{i,y}^{Border}$ | € | Total cost of the commodity i in year y. |
| $C_{i,y}^{Fix}$ | € | Fix costs of the commodity i in year y depending on the capacities installed. |
| $C_{i,y}^{Var}$ | € | Variable costs of the commodity i in year y depending on the energy produced. |
| $CAP_{n,r,c,d,i,y}^{RES}$ | MW | Capacity for renewable technologies r in country n |
| $CAP_{n,r,c,d,i,y}^{Batt}$ | MW | Capacity of battery storage |
| $CAP_{n,r,c,d,i,y}^{El}$ | MW | Capacity of electrolysers |
| $CAP_{n,r,c,d,i,y}^{Conv}$ | MW | Capacity of conversion plant |
| $Q_{n,r,c,d,i,y,h}^{RES}$ | MWh | Electricity production |
| $Q_{n,r,c,d,i,y,h}^{Conv}$ | MWh | Commodity production |
| $SOC_{n,r,c,d,i,y,h}^{Batt}$ | MWh | State of charge |
| $Q_{n,r,c,d,i,y,h}^{Batt,out}$ | MWh | Electricity output from battery |
| $Q_{n,r,c,d,i,y,h}^{Batt,in}$ | MWh | Electricity input from battery |

### 3.2.3 Model Formulations

This chapter outlines the model formulations used to perform in total six linear optimizations (three commodities multiplied by two years: 2030 and 2040).

The objective function in Eq. (1) minimizes the total cost $TC_{i,y}^{Border}$ of each commodity $i$ in year $y$, which are the fixed costs $C_{i,y}^{Fix}$ plus variable costs $C_{i,y}^{Var}$ summed over countries $n$, renewables $r$, resource classes[2] $c$, and distances to shore $d$.

$$Min\ TC_{i,y}^{Border} = \sum_{n,r,c,d} \left( C_{i,y,n,r,c,d}^{Fix} + C_{i,y,n,r,c,d}^{Var} \right) \qquad \forall i,y \qquad (1)$$

The fixed costs $C_{i,y}^{Fix}$ specified in Eq. (2) are the product of unit-specific investment costs for renewable energy sources $c_{n,r,y}^{RES}$, batteries $c_{n,y}^{Batt}$, electrolysers $c_{n,y}^{El}$, and conversion infrastructure $c_{i,y}^{Conv}$, multiplied with their respective, endogenously determined capacities $CAP$.

Full costs are annualized capital expenditures and operational expenditures. All preliminary calculations associated to the parameters can be found in the supplementary data.

$$C_{i,y,n,r,c,d}^{Fix} = c_{n,r,y}^{RES} \cdot CAP_{n,r,c,d,i,y}^{RES} + c_{n,y}^{Batt} \cdot CAP_{n,r,c,d,i,y}^{Batt} + c_{n,y}^{El} \cdot CAP_{n,r,c,d,i,y}^{El} + c_{i,y}^{Conv} \cdot CAP_{n,r,c,d,i,y}^{Conv} \qquad \forall i,y \qquad (2)$$

Variable costs $C_{i,y}^{Var}$ in Eq. (3) sum up the unit specific costs for desalination $c^{Desal}$ and transport costs within the country of origin, $c_{n,i,y}^{Tr,inl}$, multiplied by the quantity of hydrogen produced $Q_{n,r,c,d,i,y,h}^{Conv}/\eta_y^{Conv,new}$ [3], as well as unit specific cost for international transport ($c_{n,i,y}^{Tr,int}$) and storage costs $c_i^{Tank}$, multiplied by the quantity of the commodity $Q_{n,r,c,d,i,y,h}^{Conv}$ across countries $n$, renewables $r$, resource classes $c$, distances to shore $d$, and hours $h$.

$$C_{i,y,n,r,c,d}^{Var} = \sum_h \left( \left( c^{Desal} + c_{n,i,y}^{Tr,inl} \right) \cdot Q_{n,r,c,d,i,y,h}^{Conv}/\eta_{i,y}^{Conv,new} + \left( c_{n,i,y}^{Tr,int} + c_i^{Tank} \right) \cdot Q_{n,r,c,d,i,y,h}^{Conv} \right) \qquad \forall i,y \qquad (3)$$

The energy balance in Eq. (4) ensures that the summed-up quantity of each commodity $i$ in year $y$ across countries $n$, renewables $r$, resource classes $c$, distances to shore $d$, and hours $h$ is equal to the European demand in year $y$.

$$\sum_{n,r,c,d,h} Q_{n,r,c,d,i,y,h}^{Conv} = dem_y \qquad \forall i,y \qquad (4)$$

The commodity production $Q_{n,r,c,d,i,y,h}^{Conv}$ in Eq. (5) is determined by the quantity of renewable energy $Q_{n,r,c,d,i,y,h}^{RES}$, adjusted by the charging $Q_{n,r,c,d,i,y,h}^{Batt,In}$ and discharging $Q_{n,r,c,d,i,y,h}^{Batt,Out}$ of batteries, and further multiplied by the efficiencies of electrolysis $\eta_y^{El}$ and conversion processes $\eta_{i,y}^{Conv,new}$.

---

[2] Resource classes are groupings of renewable capacity factors by country, distance to shore, and technology (see Assumptions for Commodity Production).
[3] $Q_{n,r,c,d,i,y,h}^{Conv}$ is expressed in MWh of the final commodity, the factor $1/\eta_y^{Conv,new}$ introduced to convert from commodity units back to the equivalent hydrogen units.

$$Q^{Conv}_{n,r,c,d,i,y,h} = \left(Q^{RES}_{n,r,c,d,i,y,h} - Q^{Batt,In}_{n,r,c,d,i,y,h} + Q^{Batt,Out}_{n,r,c,d,i,y,h}\right) \times \eta^{El}_y \times \eta^{Conv,new}_{i,y} \qquad \forall n,r,c,d,y,i,h \quad (5)$$

The production constraints ensure that in Eq. (6) and (7) the renewable electricity production $Q^{RES}_{n,r,c,d,i,y,h}$ does not exceed installed capacities $CAP^{RES}_{n,r,c,d,i,y}$ and their capacity factors $cf^{RES}_{n,r,c,d,h}$, which are also constrained by the potential renewable capacity $pot^{RES}_{n,r,c,d}$ and that in Eq. (8) and (9) the commodity production does not exceed the installed capacities for electrolysis $CAP^{El}_{n,r,c,d,i,y}$ and conversion $CAP^{Conv}_{n,r,c,d,i,y}$, accounting for the efficiencies of these processes.

$$Q^{RES}_{n,r,c,d,i,y,h} \leq CAP^{RES}_{n,r,c,d,i,y} \times cf^{RES}_{n,r,c,d,h} \qquad \forall n,r,c,d,y,i,h \quad (6)$$

$$CAP^{RES}_{n,r,c,d,i,y} \leq pot^{RES}_{n,r,c,d} \qquad \forall n,r,c,d,y,i \quad (7)$$

$$Q^{Conv}_{n,r,c,d,i,y,h}/\eta^{Conv,new}_{i,y} \leq CAP^{El}_{n,r,c,d,i,y} \qquad \forall n,r,c,d,y,i,h \quad (8)$$

$$Q^{Conv}_{n,r,c,d,i,y,h}/\eta^{Conv}_i \leq CAP^{Conv}_{n,r,c,d,i,y} \qquad \forall n,r,c,d,y,i,h \quad (9)$$

The battery storage equations (10) to (14) manage the state of charge $SOC^{Batt}_{n,r,c,d,i,y,h=1}$ and the limits on battery capacity $CAP^{Batt}_{n,r,c,d,i,y}$, ensuring that charging $Q^{Batt,In}_{n,r,c,d,i,y,h}$ and discharging $Q^{Batt,Out}_{n,r,c,d,i,y,h}$ processes adhere to capacity and efficiency constraints. The energy-to-storage ratio, which controls how much energy the battery can store relative to its capacity, is shown in Eq. (13).

$$SOC^{Batt}_{n,r,c,d,i,y,h=1} = 0 \qquad \forall n,r,c,d,i,y,h \quad (10)$$

$$Q^{Batt,In}_{n,r,c,d,i,y,h} \leq CAP^{Batt}_{n,r,c,d,i,y} \qquad \forall n,r,c,d,i,y,h \quad (11)$$

$$Q^{Batt,Out}_{n,r,c,d,i,y,h} \leq CAP^{Batt}_{n,r,c,d,i,y} \qquad \forall n,r,c,d,i,y,h \quad (12)$$

$$SOC^{Batt}_{n,r,c,d,i,y,h} \leq CAP^{Batt}_{n,r,c,d,i,y} * 6 \qquad \forall n,r,c,d,i,y,h \quad (13)$$

$$SOC^{Batt}_{n,r,c,d,i,y,h>1} = SOC^{Batt}_{n,r,c,d,i,y,h-1} + Q^{Batt,In}_{n,r,c,d,i,y,h} \times \eta^{Batt} - Q^{Batt,Out}_{n,r,c,d,i,y,h}/\eta^{Batt} \qquad \forall n,r,c,d,i,y,h \quad (14)$$

## 3.3 Border-to-Consumer Model

This section introduces the model optimizing the supply costs of commodities from national border to consumer. It begins with an overview of the model, followed by the introduction of the nomenclature, and concludes with a description of the model formulations.

### 3.3.1 Model Overview

The Border-to-Consumer model minimizes supply costs from the import border to consumer sites, integrating transport, storage, and conversion costs.

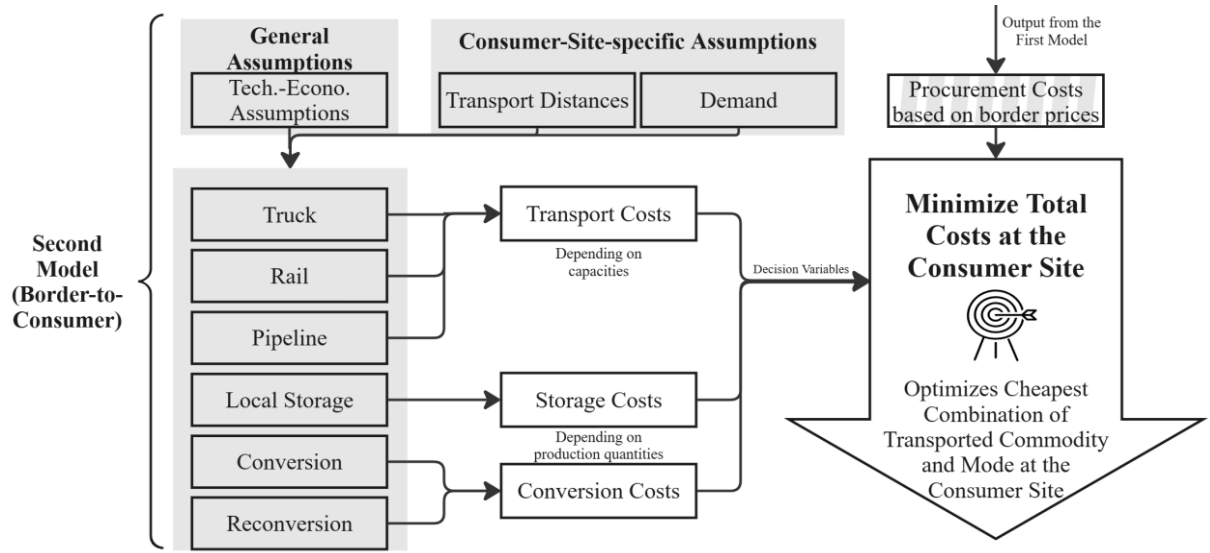

**Figure 3 Methodology for computing the supply costs of commodities at the consumer site.**

Figure 3 provides an overview of the second model, which calculates Border-to-Consumer supply costs by matching each consumer site's demand and distance with the least cost transport, storage, and conversion options. Procurement costs depend on the quantity of the imported commodity purchased at the border (as determined by the first model), while transport costs vary by mode (truck, rail, or pipeline) and the endogenously installed capacities required to move these volumes. Additional storage and conversion costs arise depending on the product and the consumer's needs, for instance, pipelines do not require storage, whereas transporting by truck or rail typically does, and conversion steps (such as cracking ammonia to hydrogen) incur further expenses. Through an optimization procedure, the model minimizes total supply costs for each consumer site by choosing the cheapest combination of commodity (e.g., ammonia or hydrogen) and transport mode, thereby yielding the minimized supply price at the consumer site.

### 3.3.2 Nomenclature

Note that we use capital letters for variables and small letters for sets and parameters.

| Abbreviation | Unit | Description |
|---|---|---|
| *Sets* | | |
| $i$ | | Commodity |
| $y$ | | Years |
| $k$ | | Transport mode |
| $j$ | | Consumer site |
| *Parameters* | | |
| $p_{i,y}$ | €/MWh | Price of commodity $i$ at the border in year $y$. |
| $c_i^{Stor}$ | €/MWh | Storage cost per unit of commodity i. |
| $c_{i,y,j}^{Conv}$ | €/MWh | Conversion or reconversion cost per unit of commodity i at site j in year y. |
| $c_{i,k,j}^{T}$ | €/unit | Transport cost per unit of commodity i via mode k to site j. |
| $dem_{y,j}$ | MWh | Demand at consumer site j in year y. |
| $cap_{i,k,j}^{Eff}$ | MWh/unit | Effective yearly transport capacity per unit for commodity i via mode k. |
| $\eta_{i,k,j}^{Cons}$ | $MW_i/MW_i$ | Supply chain efficiency for commodity i at site j using mode k. |
| $v_{i,k}$ | | Matrix indicating valid combinations of commodity i and mode k. |
| *Variables* | | |
| $TC_{y,j}^{Cons}$ | € | Total cost at site j in year y. |
| $C_{i,y,k,j}^{T}$ | € | Transport cost of commodity i via mode k to site j in year y. |
| $C_{i,y,k,j}^{SC}$ | € | Storage and conversion costs for commodity i at site j. |
| $Q_{i,y,k,j}^{P}$ | MWh | Quantity procured of commodity i via mode k to site j in year y. |
| $Q_{i,y,k,j}^{T}$ | MWh | Quantity transported of commodity i via mode k to site j in year y. |
| $I_{i,y,k,j}^{T}$ | | Number of transport units for commodity i via mode k to site j. |

### 3.3.3 Model Formulations

The model, formulated as a Mixed-Integer Programming problem, minimizes the total cost of procuring commodities at each consumer site in year *y*, considering procurement, transportation, storage, and conversion costs.

The objective function in Eq. (15) minimizes the total costs $TC_{y,j}^{Cons}$ for each year *y* and consumer site *j* by summing over all commodities (*i*) the costs of commodity procurement at the border ($Q_{i,y,j}^{P} \cdot p_{y,i}$) plus transportation costs within the country ($C_{i,y,k,j}^{T}$), as well as storage and conversion costs at the consumer site ($C_{i,y,k,j}^{SC}$), with the latter two summed over transport modes (*k*).

$$Min \; TC_{y,j}^{Cons} = \sum_{i} Q_{i,y,j}^{P} \cdot p_{i,y} + \sum_{i,k} \left( C_{i,y,k,j}^{T} + C_{i,y,k,j}^{SC} \right) \qquad \forall \; y,j \qquad (15)$$

The **first cost component**, $Q_{i,y,j}^{P} \cdot p_{i,y}$, is straightforward: A desired quantity of an energy carrier is procured at the border at a price of $p_{i,y}$, with the price being the result of the first model and thus exogenous here in Eq. (15). The **second component** is more complex. Eq. (16) explains transportation costs, which are the product of procured transport units $I_{i,y,k,j}^{T}$, which is a unitless integer variable, and the cost per transport unit.

$$C_{i,y,k,j}^{T} = I_{i,y,k,j}^{T} \cdot c_{i,y,k,j}^{T} \qquad \forall\, i, y, k, j \quad (16)$$

The amount of procured transport units $I_{i,y,k,j}^{T}$, multiplied with their capacity $cap_{i,k,j}^{eff}$, limits the total amount of commodities $Q_{i,y,k,j}^{T}$ that can be transported from border to consumption site.

$$Q_{i,y,k,j}^{T} \leq I_{i,y,k,j}^{T} \cdot cap_{i,k,j}^{Eff} \qquad \forall\, i, y, k, j \quad (17)$$

Further, the total amount of a commodity that is transported through the various transport modes needs to be procured at the border:

$$\sum_{k} Q_{i,y,k,j}^{T} \leq Q_{i,y,j}^{P} \qquad \forall\, y, j, i \quad (18)$$

The **last cost component**, storage and conversion costs at the consumer site ($C_{i,y,k,j}^{SC}$), is calculated as a linear function of the transported quantity in Eq. (19). The parameter $c_{i,k}^{stor}$ reflects that storage costs vary by the transported commodity (e.g. commodities being transported by trucks and rail require to be stored on site while pipeline-based commodities do not). The parameter $c_{i,y,j}^{conv}$ reflects costs for commodity conversion if the desired commodity of the consumer is different from the transported commodity (e.g., converting gaseous hydrogen to ammonia when consumption site requires ammonia).

$$C_{i,y,k,j}^{SC} = \left(c_{i,k}^{Stor} + c_{i,y,j}^{Conv}\right) \cdot Q_{i,y,k,j}^{T} \qquad \forall\, i, y, k, j \quad (19)$$

The **energy balance** constraint Eq. (20) is the last major constraint we would like to highlight. It ensures that the transported quantity across all commodities and transport modes, multiplied by each consumer site's supply chain efficiency $\eta_{i,k,j}^{Cons}$, meets or exceeds the demand $dem_{y,j}$ of a consumer site $j$ in each year $y$. The matrix $v_{i,k}$ specifies distribution modes[4], which are valid pairings of commodity $i$ and transport mode $k$ (see Figure 1, with entries of 1 for valid combinations and 0 otherwise). Essentially, the matrix of valid distribution modes helps to shorten notation.

---

[4] Distribution modes are valid combinations of commodity and transport mode. We consider six such distribution modes for inland transportation: ammonia by truck, rail, and pipeline; liquid hydrogen by truck and rail, and gaseous hydrogen by pipeline (see Figure 1). In the Border-to-Consumer supply chain, ammonia consumers are parameterized to either (a) procure ammonia via truck, rail, or pipeline and store it locally at the consumer site, or (b) transport liquid hydrogen via truck or rail, and convert it to ammonia through regasification and the Haber-Bosch process at the consumption site, or (c) procure gaseous hydrogen, transport it through domestic pipelines, and convert to ammonia via the Haber-Bosch process at the consumption site. For hydrogen consumers, ammonia can be procured through the same three distribution modes (truck, rail, pipeline), stored locally, and then cracked to produce gaseous hydrogen. Alternatively, liquid hydrogen is transported by truck or rail, stored, and regasified into gaseous hydrogen. Lastly, gaseous hydrogen can be delivered directly by pipeline.

$$\sum_{i,k} v_{i,k} \cdot Q^T_{y,j,i,k} \cdot \eta^{Cons}_{i,k,j} \geq dem_{y,j} \qquad \forall\, y,j \qquad (20)$$

# 4 ASSUMPTIONS AND DATA

This section provides insights into data and assumptions of the Well-to-Border model (see chapter 3.2) and the Border-to-Consumer model (see chapter 3.3).

## 4.1 Well-to-Border Assumptions

The following chapter presents the input data for the Well-to-Border model. Therefore, we describe the assumptions related to (i) production and (ii) transportation as well as (iii) the European demand needed for the energy balance.

### 4.1.1 Assumptions for Commodity Production

This subsection outlines the key assumptions for commodity production in the export countries, first detailing how renewable profiles and potentials are determined and then describing the techno-economic parameters. In total, ten export countries are selected (Algeria, Australia, Chile, Morocco, Norway, Oman, Saudi Arabia, Tunisia, United Arab Emirates, and United States of America) to capture diverse renewable potentials, distances, and bilateral trade contexts. While Morocco, Norway and Oman, and lend themselves to pipeline or short-range shipping, more distant exporters such as Australia and Chile benefit from abundant renewable resources but face higher transport costs. The United States and the United Arab Emirates were included to increase scenario realism, given their established agreements with Germany [46].

Our study applies the Atlite software to determine country-specific renewable profiles and potentials, taking into account various criteria including protected areas, land cover, and technical restrictions [47]. Weather-dependent renewable energy variability is captured using 2018 meteorological data. Following Sens et al. [27], we limiting the technical potentials using an availability factor to reflect socio-economic realities. In line with the methodology outlined by Brändle et al. [48], we categorize these profiles into four technologies: Photovoltaic, onshore wind, and offshore wind (which is split in two technologies due to cost differences, namely shallow and deep water). All four technologies are subdivided into five quality classes[5] according to their capacity factors. Additionally, the profiles are grouped into regional categories based on their proximity to the coast, segmented at 250 km intervals, to facilitate later assessments of inland transportation efforts [49].

---

[5] First, capacity factors are calculated for each location and technology. Next, the lowest and highest values among those factors are identified. Finally, the range from lowest to highest is divided into five equal segments, and each segment constitutes one resource class.

The techno-economic parameters related to renewables, batteries, electrolysis, and conversion processes are detailed in Table 1. All costs are reported in €$_{2023}$ and all calorific units correspond to the lower heating value. Consistent with recommendations by Genge et al. [19], we have focused on sensitive parameters such as the Weighted Average Cost of Capital[6] and conducted a comprehensive literature review on the capital expenditures of production processes [21], [27], [50], [51], [52], [53]. The review includes technologies such as photovoltaic, onshore wind, offshore wind (both shallow and deep water), Polymer Electrolyte Membrane electrolysis, the Haber-Bosch process, and processes for liquefaction, as well as cracking and regasification as reconversion. Capital expenditure values presented in this study represent the mean values derived from our literature review. Operational expenditures, efficiencies, power demands, and lifetimes of processes along with their scaling factors are incorporated based on findings from Moritz et al. [21]. Specifically, the capital expenditure for the Haber-Bosch process includes the costs associated with the air separation unit necessary for extracting nitrogen, with detailed costs and electricity demands sourced from Dry [54] and Bazzanella and Ausfelder [55].

Finally, Table 2 summarizes the techno-economic parameters for energy storage. Based on Sens et al. [27], costs for battery technologies are noted to vary over time, reflecting market technological changes, whereas costs for other storage solutions like salt caverns, ammonia, and liquid hydrogen tanks remain stable. These options are assessed as large-scale facilities.

**Table 1 Techno-economic assumptions for electricity generation, hydrogen production and (re)conversion. Capital expenditures of renewables have been adjusted to reflect country-specific costs based on an index that accounts for the variability in local labor costs, material availability, and material pricing** [51].

| Technology | | 2030 | 2040 | Unit | Source |
| --- | --- | --- | --- | --- | --- |
| **Power Generation** | | | | | |
| PV | capex[7] | 425-511 | 368-433 | €/kW | Lit. Rev. |
| | opex | 2.6 | 2.7 | % of capex | [27] |
| | lifetime | 30 | 30 | years | [27] |
| | | | | | |
| Wind Onshore | capex[7] | 987-1,207 | 917-1,135 | €/kW | Lit. Rev. |
| | opex | 2.5 | 2.4 | % of capex | [27] |
| | lifetime | 30 | 30 | years | [27] |
| | | | | | |
| Wind Offshore Shallow-Water | capex[7] | 1,849-1,902 | 1,620-1,710 | €/kW | Lit. Rev. |
| | opex | 2.8 | 2.7 | % of capex | [27] |
| | lifetime | 30 | 30 | years | [27] |
| | | | | | |
| Wind Offshore Deep-Water | capex[7] | 2,631-2,667 | 2,300-2,357 | €/kW | Lit. Rev. |
| | opex | 2.8 | 2.7 | % of capex | [27] |

---

[6] WACC
[7] Capex of renewables is given as a range of country-specific values from the literature review. All values are listed in the supplementary data.

|  |  | | | Unit | Source |
|---|---|---:|---:|---|---|
|  | lifetime | 30 | 30 | years | [27] |
| **Hydrogen Production** | | | | | |
| Electrolysis | capex | 801 | 640 | €/kW | Lit. Rev. |
|  | opex | 3.5 | 3.2 | % of capex | [27] |
|  | eff | 0.67 | 0.69 | kWel/kWGH2 | [27] |
|  | lifetime | 20 | 23 | years | [27] |
| **Conversion** | | | | | |
| Haber-Bosch | capex | 1,101 | 963 | €/kW | Lit. Rev. |
|  | opex | 4 | 4 | % of capex | [21] |
|  | Power demand | 0.29 | 0.29 | kWel/kWNH3 | [21] |
|  | eff | 0.85 | 0.85 | kWNH3/ kWGH2 | [21] |
|  | lifetime | 25 | 25 | years | [21] |
| Liquefication | capex | 1,062 | 584 | €/kW | Lit. Rev. |
|  | opex | 4 | 4 | % of capex | [21] |
|  | Power demand | 0.20 | 0.20 | kWel/kWLH2 | [56] |
|  | eff | 0.9 | 0.9 | kWLH2/kWGH2 | [56] |
|  | lifetime | 30 | 30 | years | [21] |
| **Reconversion** | | | | | |
| Cracking | capex | 764 | 575 | €/kW | Lit. Rev. |
|  | opex | 4 | 4 | % of capex | [21] |
|  | Power demand | 0.05 | 0.05 | kWel/kWNH3 | [56] |
|  | eff | 0.78 | 0.78 | kWNH3/kWGH2 | [56] |
|  | lifetime | 25 | 25 | years | [57] |
| Regasification | capex | 812 | 593 | €/kW | Lit. Rev. |
|  | opex | 4 | 4 | % of capex | [21] |
|  | Power demand | 0.01 | 0.01 | kWel/kWLH2 | [21] |
|  | eff | 0.9 | 0.9 | kWLH2/kWNH3 | [21] |
|  | lifetime | 30 | 30 | years | [21] |

**Table 2 Storage technologies.**

|  | Battery 2030 | Battery 2040 | Salt Cavern GH2 | Tank NH3 | Tank LH2 | Unit | Source |
|---|---:|---:|---:|---:|---:|---|---|
| capex | 190 | 143 | 1,465 | 414 | 1,051 | €/MWh | [11] |
| opex | 2 | 3 | 2 | 2 | 2 | % of capex | [11] |
| eff | 0.98 | 0.98 | 1 | 1 | 1 | % | [11] |
| lifetime | 15 | 15 | 30 | 30 | 30 | years | [11] |

### 4.1.2 Transport from Well-to-Border

This subsection presents the assumptions for the transport from well to border in the first model.[8] First, hydrogen pipelines within the exporting country connect production facilities to export nodes, with their techno-economic parameters outlined in Table 3. For international transport, nearby exporters may also use (large) hydrogen pipelines to Europe, whereas more distant exporters rely on shipping (Table 4). In the case of liquid hydrogen, any boil-off losses aboard ships are assumed to be used as fuel, thereby improving overall energy efficiency.

Table 3 Hydrogen Pipeline Transport

| Pipeline | International GH2 | Domestic GH2 | Unit | Source |
|---|---|---|---|---|
| capex | 6,982,749 | 1,146,105 | €/km | [27], [29] |
| opex | 0.05 | 0.05 | % capex/a | [27], [29] |
| lifetime | 40 | 40 | a | [27], [29] |
| cf | 0.90 | 0.90 | % | [27], [29] |
| Hydrogen transport | 172,998,270 | 9,266,574 | MWh/a | [27], [29] |
| Electricity demand | 0.00002 | 0.00002 | MWhel/(MWh 1km) | [27], [29] |
| Average electricity price | 53 | 53 | €/MWhel | [51] |
| Hydrogen loss | 0 | 0 | kgH2/(kgH2 100km) | [27], [29] |

Table 4 Seaborne Transport

| Shipping | LH2 | NH3 | Unit | Source |
|---|---|---|---|---|
| capex | 410,687,496 | 83,835,432 | €/ship | [27] |
| opex | 0.04 | 0.04 | % of capex | [27] |
| lifetime | 25 | 25 | a | [27] |
| operating costs | 607 | 607 | €/h | [27] |
| cf | 8,000 | 8,000 | h/a | [27] |
| velocity | 33 | 30 | km/h | [27] |
| fuel demand | 0 | 0.69 | MWh/km | [27] |
| fuel cost | 170 | 170 | €/MWh | [27] |
| payload | 366,663 | 311,664 | MWh/ship | [27] |
| de/loading time | 54 | 54 | h/load | [27] |
| hydrogen loss | 0 | 0 | %/h | [27] |
| hydrogen flash | 0.01 | 0 | %/load | [27] |

---

[8] Following Lux et al.[49], we use Germany, given its central position in Europe, as a representative location for import price calculations.

### 4.1.3 European Demand

European demand levels of 306 TWh in 2030 and 861 TWh in 2040 set the benchmark for evaluating import prices of each commodity at the European border. The model incorporates the mean demand scenario following the meta-analysis by Riemer et al. [58]. These demand projections are normalized to lower heating values and serve as a benchmark for each commodity.

The justification for using hydrogen demand as the benchmark for hydrogen-based commodities lies in their shared reliance on hydrogen, which creates competition for the same renewable energy resources. By applying hydrogen demand as the baseline, the model provides a consistent framework to accurately map total demand across all these interconnected commodities. This approach addresses uncertainties regarding future demand for specific hydrogen-based commodities and ensures flexibility, allowing the assessment to remain applicable.

## 4.2 Border-to-Consumer Assumptions

This section parameterizes the Border-to-Consumer model focusing on domestic transport and consumer sites. Storage and conversion technologies align with the first model and are detailed in Table 1 and Table 2. Although the methodology itself is globally applicable, the specific data and assumptions employed here are tailored to a European context.

### 4.2.1 Transport from Border-to-Consumer

Data and assumptions for truck and rail transport of ammonia and liquid hydrogen are outlined in Table 5, while ammonia and hydrogen pipeline details are in Table 6. Road transport uses hydrogen fuel cell-powered tractors to support climate-neutral goals, and rail transport uses an electrified TRAXX F140 MS2 locomotive [59], [60]. Transport parameters remain stable over time [27].

**Table 5 Road and Rail Transport**

|  | Truck Trailer NH3 | Truck Trailer LH2 | Tractor | Rail Trailer NH3 | Rail Trailer LH2 | Tractor | Unit | Source |
|---|---|---|---|---|---|---|---|---|
| capex | 212,242 | 965,699 | 201,630 | 212,242 | 965,699 | 2,980,900 | €/unit | [27] |
| cap | 87 | 133 |  | 87 | 133 |  | MWh | [27], [59] |
| lifetime | 12 | 12 | 5 | 12 | 12 |  | year | [27] |
| through boil-off | 0 | 0.5 |  |  | 0.5 |  | % of throughput | [59] |
| through | 1 | 1 |  | 1 | 1 |  | % | [59] |
| boil-off | 0.0 | 0.3 |  |  | 0.3 |  | %/d | [59] |
| speed |  |  | 50 |  |  | 50 | km/h | [59] |
| time_load |  |  | 1.5 |  |  | 1.5 | h/load | [59] |
| lab_driver |  |  | 38 |  |  |  | €/h | [59] |
| fuel_econ |  |  | 0.0023 |  |  |  | km/L | [27] |
| fuel_cost_2030 |  |  | 140 |  |  |  | €/L | [59] |

| | | | €/L | [59] |
| --- | --- | --- | --- | --- |
| fuel_cost_2040 | 70 | | | |
| freight rate | | 4.75 | €/km | [59] |

**Table 6 Domestic Pipeline Transport**

| Pipeline | Domestic GH2 | Domestic NH3 | Unit | Source |
| --- | --- | --- | --- | --- |
| capex | 1,146,105 | 502,084 | €/km | [27], [29] |
| opex | 0.05 | 0.03 | % capex/a | [27], [29] |
| lifetime | 40 | 40 | a | [27], [29] |
| cf | 0.90 | individual | % | [27], [29] |
| H2_transport | 9,266,574 | 212,242 | MWh/a | [27], [29] |
| Electricity_demand | 0 | 0 | MWhel/(MWh 1km) | [27], [29] |
| ElectricityPrice_Avg | 53 | 53 | €/MWhel | [51] |
| H2_loss | 0 | 0 | kgH2/(kgH2 100km) | [27], [29] |

### 4.2.2 Consumer Sites

Consumer sites vary by distance from the border and the size of their demand (in our model set-up, consumption sites demand a positive amount of either hydrogen or ammonia). We employ the second model to analyse both generic and empirical consumer sites.

The generic sites, 100 in total, are organized into a matrix of ten annual demand levels (ranging from 10 GWh to 10,000 GWh) by ten distances, starting at 5 km (representing a consumer located in close proximity to the import hub) and then increasing from 100 km up to 900 km in 100 km increments.

In addition, we analyse fourteen empirical consumer sites within Germany, chosen due to robust data availability [61] and their usage of hydrogen and ammonia. Ammonia consumers are represented by 4 basic chemical industry sites and one ship fuelling station. The demands of the basic chemical industry (fertilizer and chemical precursor industry) are based on Neuwirth et al. [61]. Their output is expected to remain constant until 2050. Ammonia can also be used in the shipping industry [62], [63]. While not yet existing, we examine its feasibility for inland shipping consumers with one exemplary fuelling station along the River Rhine, located in Ludwigshafen. The demand for the fuelling station is estimated by integrating various factors, including fuel demand, transport volume per ship, distance from Rotterdam Port, daily ship traffic, and detour considerations. Further, we include nine hydrogen consumers, subgrouped into three types. First, to estimate the consumption of three hydrogen-fired power plants we have collected information on the project announcements [64], [65], [66] and average yearly capacity factors [67]. Second, the demand of two hydrogen fuelling stations for heavy-duty transport is derived from Reuß et al. [68]. Finally, the amount of hydrogen feedstock for two chemical industry sites, one steel plant and one refinery, are taken from Neuwirth et al. [61]. Each empirical consumer site is identified through their demands and transport mode-specific distances (Table 7). Distances are measured between import node and consumer site. While truck and rail distance are determined using actual routing via OpenStreetMap, pipeline distances following European Hydrogen Backbone development plans [69]. Whereas the empirical sites distinguish between two import nodes, one terminal (ammonia and liquid hydrogen) and one pipeline (gaseous hydrogen), in the generic analysis, a single import node is used for all commodities.

Table 7 Consumer-specific demands and distances.

| Consumer sites | Desired Product | Demand [GWh] 2030 | Demand [GWh] 2040 | Distance [km] Truck | Distance [km] Rail | Distance [km] GH2 Pipe | Distance [km] NH3 Pipe |
|---|---|---|---|---|---|---|---|
| BASF | Ammonia | 3,844 | 3,844 | 656 | 678 | 153 | 685 |
| INEOS Manufacturing Deutschland GmbH | Ammonia | 1,163 | 1,163 | 500 | 508 | 290 | 490 |
| SKW Stickstoffwerke Piesteritz GmbH | Ammonia | 6,212 | 6,212 | 444 | 414 | 693 | 448 |
| YARA Brunsbüttel GmbH | Ammonia | 3,769 | 3,769 | 13 | 4 | 763 | 3 |
| Ship Bunkering Station Ludwigshafen | Ammonia | 87 | 87 | 658 | 678 | 154 | 794 |
| Fueling Station Berlin | Hydrogen | 8 | 8 | 368 | - | 823 | 437 |
| Fueling Station München | Hydrogen | 8 | 8 | 864 | - | 290 | 924 |
| Dow Europe Holding B.V. | Hydrogen | 1,730 | 9,062 | 63 | 153 | 732 | 52 |
| Basell Polyolefine GmbH Werk Wesseling | Hydrogen | 3,184 | 16,680 | 516 | 520 | 253 | 521 |
| OMV Werk Burghausen | Hydrogen | 1,486 | 7,783 | 961 | 920 | 622 | 959 |
| Salzgitter Flachstahl GmbH | Hydrogen | 9,75 | 5,107 | 302 | 291 | 295 | 573 |
| Power Plant Altbach/Deizisau (EnBW) | Hydrogen | 18 | 222 | 757 | 779 | 256 | 806 |
| Power Plant Leipzig Süd (Stadtwerke Leipzig) | Hydrogen | 7 | 37 | 487 | 561 | 630 | 507 |
| Power Plant Schwarze Pumpe (LEAG) | Hydrogen | 6 | 30 | 543 | 550 | 818 | 770 |

# 5 RESULTS & DISCUSSION

The results section is structured into two main parts: First, it examines import prices at the border for gaseous hydrogen, liquid hydrogen, and ammonia in 2030 and 2040, analysing cost trends, global merit orders, and shadow price compositions. Second, it explores overall supply costs at consumer sites, focusing on transport modes, demand scenarios, and the role of ammonia as both a direct commodity and a hydrogen carrier.

## 5.1 Import Prices at the Border

The Well-to-Border model calculates the marginal supply costs for gaseous hydrogen, liquid hydrogen, and ammonia in 2030 and 2040. Table 8 shows that gaseous hydrogen is the least expensive option, with supply costs of 99 €/MWh in 2030 and 84 €/MWh in 2040. Ammonia maintains a cost advantage over liquid hydrogen in both years, although diminishing from a 16 % difference in 2030 to 10 % in 2040. This trend aligns with literature, which predicts a gradual reduction in the price differential between ammonia and liquid hydrogen (see Literature Review). However, gaseous hydrogen's cost leadership relays on new pipeline infrastructure and is limited to the European catchment area. In contrast, ammonia offers a competitive alternative for long-distance transport and storage. Its narrowing cost advantage over liquid hydrogen reflects ongoing technological advancements in conversion, transport and storage processes.

Table 8 Import prices at the border of three commodities in years 2030 and 2040.

| Commodity | Price [€/MWh] | |
|---|---|---|
| | 2030 | 2040 |
| Ammonia | 141 | 123 |
| Liquid Hydrogen | 167 | 137 |
| Gaseous Hydrogen | 99 | 84 |

The following chapters examine global merit orders and shadow price compositions. Additional detail on the underlying country-specific factors, including renewable resource quality and investment risk, is provided in Appendix A.

### 5.1.1 Ammonia

Initially, we analyse the import prices of ammonia at the border in 2030 and 2040.

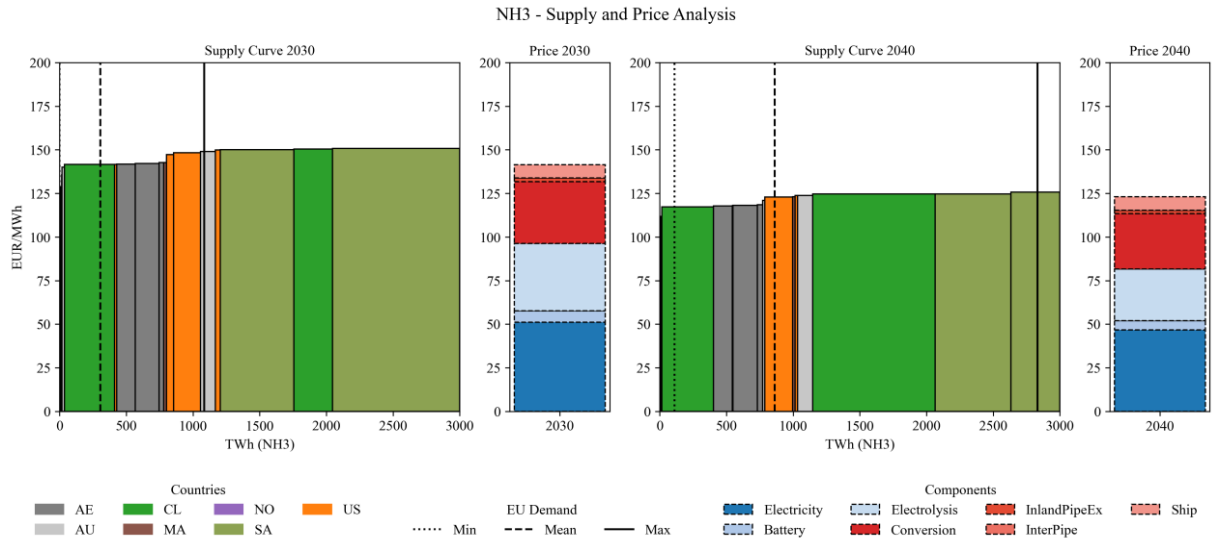

**Figure 4 illustrates global supply curves and prices of ammonia at the border for 2030 and 2040.**

Figure 4 displays global supply curves[9] for ammonia import prices in 2030 and 2040, illustrating how marginal costs evolve with increasing supply volume. EU demand thresholds (minimum, mean, and maximum) are delineated by dashed and solid vertical lines, while colour-coded bars highlight the comparative efficiency and potential of different exporting countries. The bar charts break down the price components for the mean demand scenario.

Two main insights emerge. First, the supply curves remain flat. Despite the EU demand thresholds varying by factors of 26, the impact on import price is small, increasing only by 20 %. The most cost-effective suppliers of ammonia are identified as Chile, the United Arab Emirates, and the United States. Second, ammonia import prices fall by roughly 13% over time, driven in part by declining costs for electricity (-12 %), batteries (-7 %), electrolysers (- 18 %), and conversion (-4 %). Moreover, higher consumption in 2040 encourages additional sourcing from the United Arab Emirates and the United States, further emphasizing their economic competitiveness.

### 5.1.2 Liquid Hydrogen

Next, we assess liquid hydrogen import prices at the border in 2030 and 2040. The merit order and shadow price composition are shown in Figure 5.

---

[9] For the merit order visualizations in Figure 4, Figure 5 and Figure 6, the demand parameter $dem_y$ in Eq. (4) was set to 3,000 TWh to display the full supply range of ammonia and hydrogen. This allows the visualization beyond the mean demand level. In contrast, using mean demand as the upper bound (as we do it to generate the shadow prices) would result in supply curves ending at the mean demand threshold.

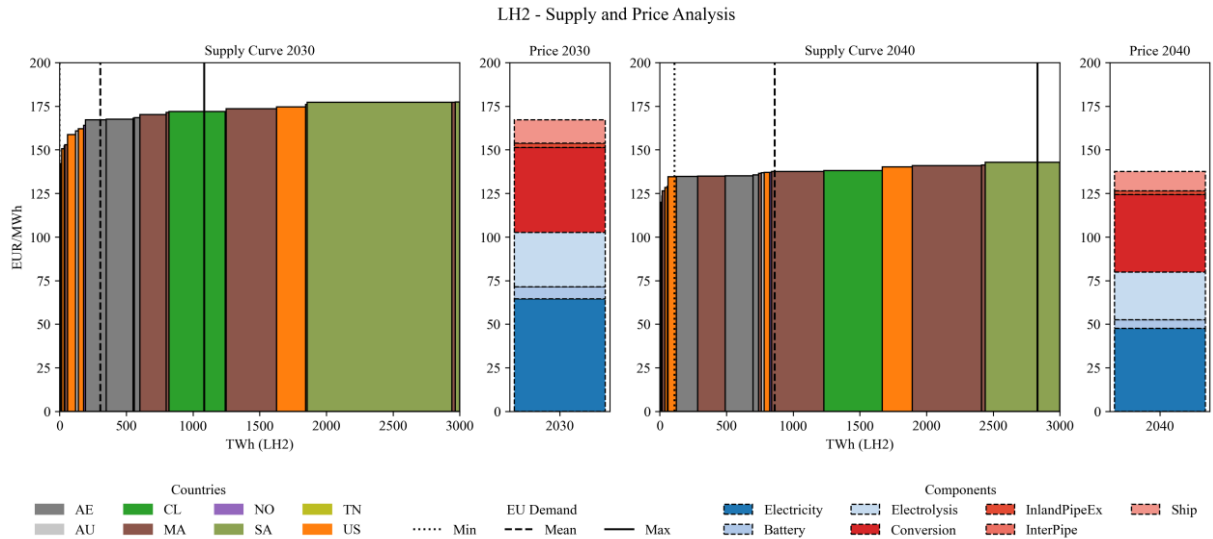

**Figure 5 illustrates global supply curves and prices of liquid hydrogen at the border for 2030 and 2040.**

Figure 5 presents the liquid hydrogen supply curve and prices. Similar to that of ammonia, it remains flat, with only modest cost increases of +6 % across different demand scenarios, underscoring a stable supply. The most cost-effective suppliers identified are Morocco, the United Arabic Emirates, and the United States. Higher shipping costs of liquid hydrogen result into selected importing countries with shorter transport distances (e.g. investing into Morocco in 2040 reduces shipping costs by 8 %). By 2040, liquid hydrogen import prices are projected to decrease by -18 %, driven by reductions in electricity (-15 %), battery (-19 %), and electrolysis (-18 %) and liquefaction costs (-15 %).

### 5.1.3 Gaseous Hydrogen

Finally, we analyse gaseous hydrogen supply costs at the border in 2030 and 2040.

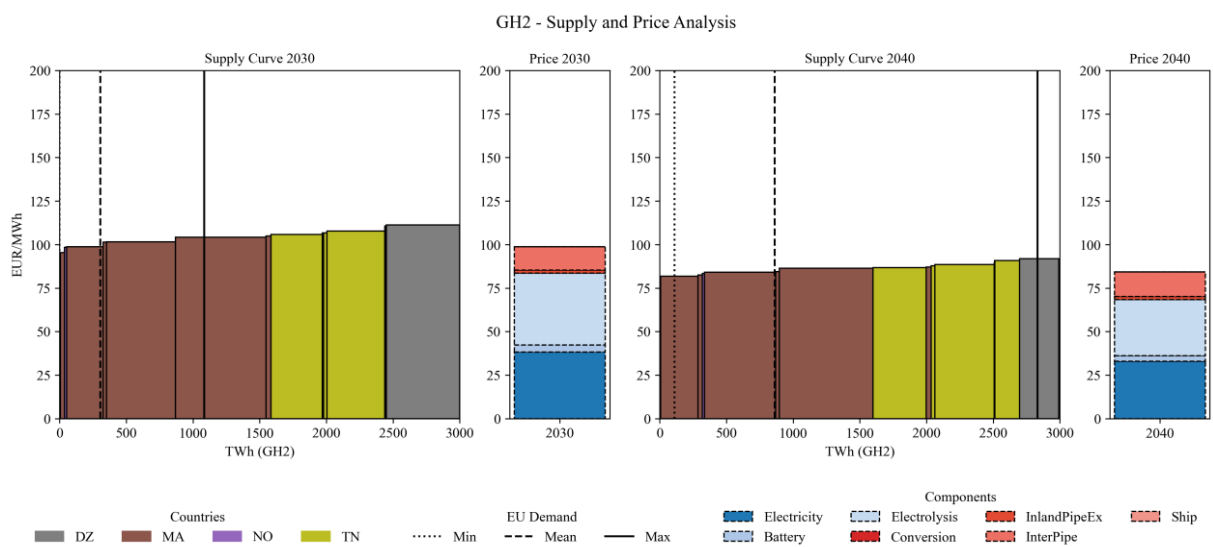

**Figure 6 presents comparative bar charts detailing the global supply costs of gaseous hydrogen at the border for the years 2030 and 2040.**

Figure 6 shows two findings on gaseous hydrogen import prices at the border for 2030 and 2040. First, the supply curve for gaseous hydrogen remains relatively flat, with minimal cost increases across varying consumption scenarios, showing only a rise of up to 11 %. Morocco identified as the most cost-effective suppliers which is able to cover entire EU demand with its high PV and onshore wind potentials. Second, there is a significant overall reduction in import prices by 2040, with a -15 % drop primarily driven by cost decreases in batteries (-10 %), electrolysis (-17 %), and electricity (-13 %). However, this decrease is offset by increased inland and international pipeline costs (+19 % or +1.78 €/MWh), as more distant production sites are utilized to meet growing demand in 2040.

## 5.2 Supply Costs at the Consumer Site

This section analyses and discusses the results of the Border-to-Consumer model, highlighting ammonia's dual role as a commodity for direct consumption and as a transport medium for hydrogen consumers. First, we demonstrate the full capabilities of the model showing the results for 100 generic sites representing eligible consumers across Europe. Following this, the model is applied to 14 empirical sites in Germany to further illustrate its practical application. A detailed examination of the Border-to-Consumer model is provided in Appendix B, where we delve into cost composition, distribution mode selection, and how supply costs change with demand and distance, particularly for hydrogen consumers in 2030.

### 5.2.1 Generic Consumer Sites

We computed the supply costs from Border-to-Consumer for 100 generic ammonia and hydrogen consumers in 2030 and 2040. The analysis optimizes total supply costs by selecting the most cost-effective imported commodity and distribution mode for each consumer site, considering varying distances and demand levels.

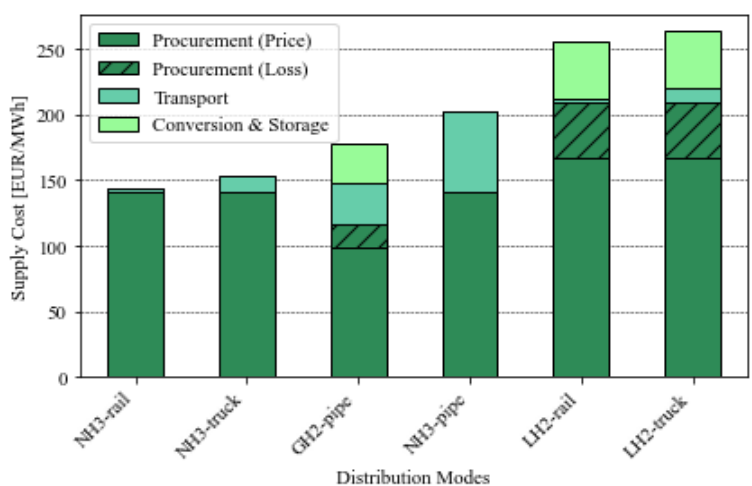

**Figure 7 Exemplary cost breakdown of one ammonia consumer with a distance of 400 km and a demand of 1,000 GWh in 2030.**

Figure 7 presents the total supply costs for a 1,000 GWh ammonia consumer located 400 km from the import hub in 2030. Across six distribution modes, its costs range from approximately 145 €/MWh to 263 €/MWh. Each mode combines a transport method (truck, rail, or pipeline) with one commodity (ammonia, liquid hydrogen, or gaseous hydrogen). For this consumption site, direct ammonia transport is the least expensive, followed by gaseous and liquid hydrogen. Rail transport is the most cost-effective mode, outperforming trucks and pipelines. Consequently, ammonia by rail is the cheapest option, while liquid hydrogen via rail or truck are most costly. Procurement expenses, influenced by conversion losses (17 €/MWh for gaseous hydrogen, 42 €/MWh for liquid hydrogen), dominate overall costs. Transport expenses range significantly, from 2 €/MWh for ammonia rail to 60 €/MWh for ammonia pipeline which is resulting from an integer investment decision and relatively low site consumption, leading to a low utilisation. Conversion and storage add roughly 31 €/MWh for gaseous hydrogen and 43 €/MWh for liquid hydrogen, with ammonia storage costs negligible at under 1 %. Further insights are detailed in Appendix B.

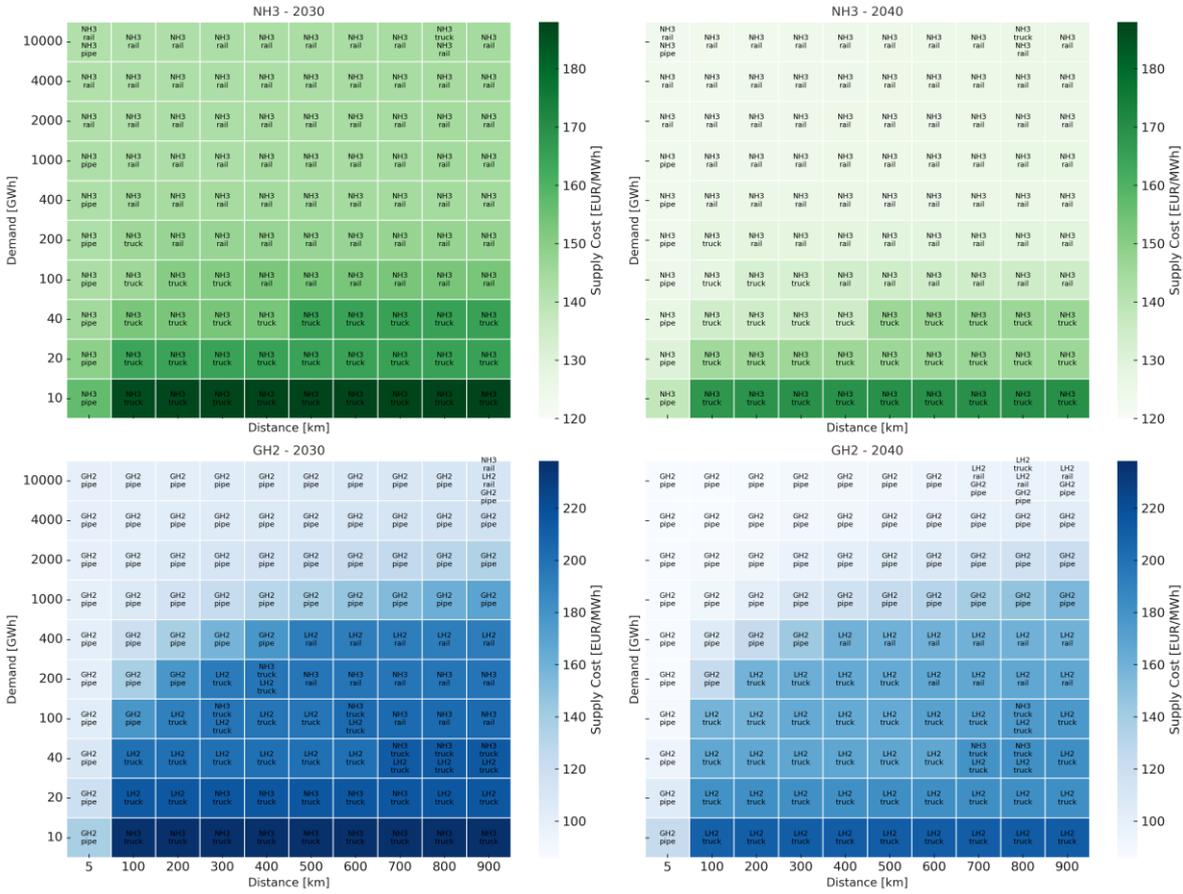

**Figure 8. Total supply costs for generic consumption sites of ammonia (green) and hydrogen (blue) in EUR/MWh (shade of color) and distribution mode (annotation).**

We have done this type of analysis for all combinations of demand and distance, and calculated the cheapest procurement strategy. The results, i.e. the cheapest distribution mode and the resulting supply cost, are depicted in Figure 8. The figure reveals significant trends regarding supply costs, distribution modes, and ammonia's role as a transport medium. A notable finding is the absence of a universal distribution mode, with optimal transport methods varying significantly based on distance and demand. Truck transport dominates in scenarios with low demand, whereas rail transport emerges as the preferred choice for medium distances and moderate demands. For gaseous hydrogen consumers facing large distances and high demand, pipeline transport is most cost-effective.

Over time, total supply costs decrease due to declining procurement, transport, and conversion expenses. Specifically, ammonia consumers experience a reduction in costs from 142-188 €/MWh in 2030 to 124-170 €/MWh by 2040. For hydrogen consumers, costs decline from 100-238 €/MWh in 2030 to 86-212 €/MWh by 2040. Generally, lower costs align with higher demand and shorter distances, whereas higher costs correlate with longer distances and lower demand. Inland transport costs can constitute up to 46 % of total costs in scenarios of low-consumption, long-distance truck transport or medium consumption via long pipeline transport, though the average is notably lower (13-14 % for hydrogen consumers and 6-7 % for ammonia consumers).

Direct ammonia consumption clearly dominates for ammonia consumers due to its cost advantages in both procurement and transport. The optimal transport mode for ammonia consumers depends on demand and distance. Truck transport is feasible for all distances up to 20 GWh demand and remains favourable for shorter distances up to 100 GWh/a. When demand exceeds two trucks' capacity, rail becomes the preferred option, handling most generic ammonia consumer sites and presumably serving the majority of national demand. Ammonia pipelines are limited to very short distances (approximately 5 km), representing consumer sites at import nodes.

For hydrogen consumers, ammonia's role as a transport commodity is significant in 2030, despite the extra reconversion costs involved. It is particularly preferred for truck-based transport at low demands (10 GWh/a) across all distances, and for rail transport at moderate demands (100-200 GWh/a) over longer distances exceeding 600 km. However, by 2040, ammonia's role diminishes significantly, driven by reduced procurement costs of liquid hydrogen, rendering it a more competitive alternative. Gaseous hydrogen pipelines emerge as the most economical solution for large hydrogen consumers (above 1,000 GWh) and even smaller consumers at short distances, primarily due to the elimination of reconversion costs.

Regarding behavioural shifts and economic considerations, ammonia remains the preferred transport commodity for ammonia consumers. However, hydrogen consumers increasingly favour direct hydrogen supply, suggesting ammonia's role as a bridging technology is limited and temporary. Investments in ammonia crackers thus pose potential risks of stranded assets, highlighting the need for cautious infrastructure planning.

## 5.2.2 Empirical Consumer Sites

The following results applying the Border-to-Consumer model to empirical consumer sites in Germany. A detailed description of the data and assumptions of the empirical sites can be found in chapter 4.2.2.

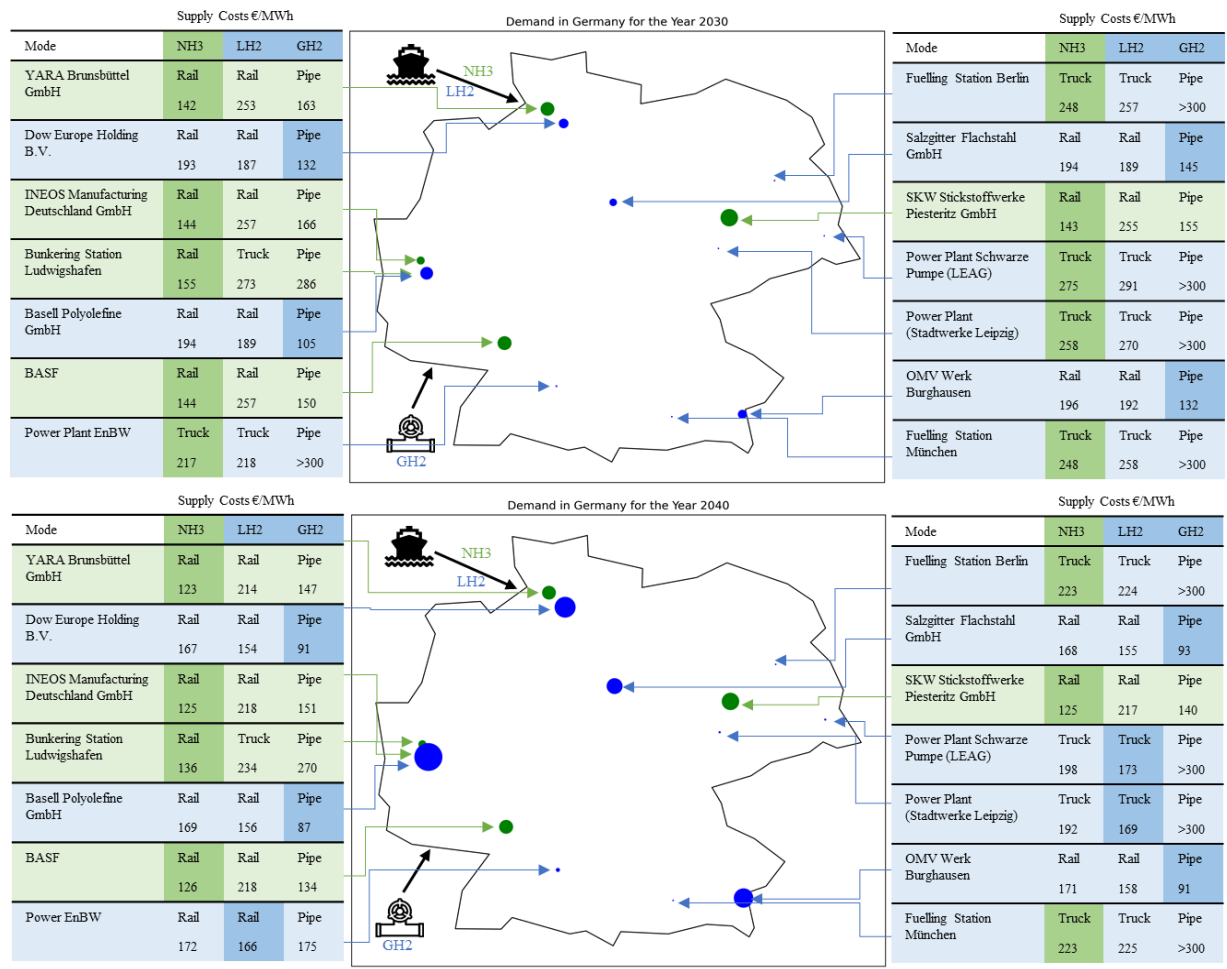

**Figure 9 displays the total supply costs at the empirical consumer site in 2030 (upper part of figure) and 2040 (lower part of figure).**

Figure 9 presents the supply costs for 14 empirical consumer sites in Germany, highlighting the varying demand (represented by dot size) and distribution modes for each commodity, with associated costs clearly indicated. The analysis identifies three main insights. In 2030, ammonia transport by truck and rail dominates distribution modes for ammonia consumers and smaller hydrogen consumers, resulting in hydrogen power plants and fuelling stations largely depending on ammonia supplies with local cracking as needed. Conversely, larger hydrogen consumers, including steel plants, refineries, and the chemical industry, benefit most from gaseous hydrogen pipelines.

By 2040, declining procurement costs for liquid hydrogen lead to significant shifts. Hydrogen power plants increase their demand and transition to liquid hydrogen supplies, whereas ammonia consumers maintain direct ammonia supply. For large-scale hydrogen consumers, pipelines remain the most cost-effective distribution mode. Consistent with trends observed at generic sites, smaller hydrogen fuelling stations and power plants temporarily rely on ammonia as a transport medium but transition to direct hydrogen use in the longer term, particularly in power plants. In scenarios lacking pipeline infrastructure in 2030, liquid hydrogen via rail would serve large hydrogen consumers.

### 5.2.3  Limitations and Potential Extensions

While comprehensive, the current study has several limitations. First, inherent uncertainties, particularly in the transition towards green energy carriers, including techno-economic developments (e.g., costs for solar power and electrolysers), infrastructure implementation (pipeline availability within and beyond Europe), and final hydrogen demand, significantly influence the results. Second, exploring additional commodities such as methanol could offer further insights, particularly regarding infrastructure requirements for the high-value chemical industry.

Moreover, this analysis centres on empirical consumer sites exclusively in Germany. Expanding the analysis to additional European locations would increase the generalizability of the findings, and a broader dataset of consumer sites would yield a more comprehensive understanding of infrastructure requirements and optimal supply strategies across diverse regions. Lastly, the role of domestic ammonia and hydrogen production has not been considered. Future research should evaluate local production's feasibility and its relationship with international imports. A comparative assessment between domestic production and imported ammonia or hydrogen would deliver valuable insights into establishing cost-effective and robust green ammonia supply chains.

## 6  CONCLUSION

This study provides a comprehensive techno-economic analysis of green ammonia supply chains, covering the entire process from electricity generation in ten exporting countries to generic and empirical consumer sites in Europe. We presented two distinct yet interconnected models (Well-to-Border and Border-to-Consumer) allowing for flexibility in integrating external assumptions. Our analysis comparatively assesses the cost-effectiveness of ammonia with liquid hydrogen, and gaseous hydrogen, revealing several critical findings.

First, empirical results indicate that final consumption largely dictates the optimal import vector. Ammonia consumers prefer direct ammonia imports, despite gaseous hydrogen being cheaper at the border. This preference is primarily driven by ammonia's cost-effectiveness in transport and direct use. Conversely, hydrogen consumers favor the hydrogen import vector, with large-scale consumers opting for pipeline-based transport for both international and domestic distribution. Smaller hydrogen consumers generally rely on liquid hydrogen transported domestically, although very limited quantities may occasionally be transported as ammonia.

At the Well-to-Border level, gaseous hydrogen is the most economical import option in both 2030 and 2040. While ammonia maintains a cost advantage over liquid hydrogen, technological advancements in liquefaction and transportation narrow this gap from 16 % in 2030 to 10 % in 2040. Competitive ammonia suppliers (primarily Morocco, the United States, and the United Arab Emirates) benefit significantly from low renewable energy costs and advanced technological readiness. Import prices are expected to decline notably by 2040 due to reductions in costs associated with electricity (- 13 %), electrolyzers (- 17 %), and conversion technologies (- 4 %).

The Border-to-Consumer analysis further clarifies that the optimal distribution mode heavily depends on consumer-specific demand and transport distance. Trucks are optimal for low demand, rail transport suits medium distances and moderate demand, and pipelines are the most cost-effective for large distances and high demand scenarios. While ammonia remains the primary transport option for ammonia consumers, its role as a hydrogen transport medium is limited and diminishes by 2040. Thus, ammonia's utility as an intermediary transport commodity for hydrogen could be temporary, and investments in ammonia cracking infrastructure may carry the risk of becoming stranded assets.

Our findings imply additional important policy implications. Given that gaseous hydrogen remains the cheapest import option, prioritizing pipeline infrastructure helps to minimize import costs, particularly for large-scale hydrogen consumers. However, ammonia still has substantial potential as an initial green hydrogen-based commodity, significantly contributing to the decarbonization of existing and future direct ammonia demands. Policymakers should recognize ammonia's strategic importance for immediate decarbonization efforts and direct ammonia applications, carefully balancing near-term infrastructure investments against the long-term risk of stranded assets in ammonia cracking facilities. Furthermore, ammonia is the more flexible import vector. First, investment costs can be ramped up ship-by-ship and second, ships can be re-routed. In contrast, hydrogen pipelines have higher lump-sum costs and very limited possibilities to re-route flows. However, as uncertainty was not explicitly addressed in our analysis, we specifically recommend such analyses for further research.

**CRediT authorship contribution statement**

**L. Genge**: Writing – original draft, Writing – review & editing, Visualization, Validation, Modelling, Methodology, Investigation, Formal analysis, Data curation, Conceptualization. **F. Müsgens**: Writing – original draft, Writing – review & editing, Supervision, Resources, Funding acquisition, Methodology, Conceptualization.

**Declaration of competing interest**

The authors declare that they have no known competing financial interests or personal relationships that could have appeared to influence the work reported in this paper.

**Acknowledgments**

LGe gratefully acknowledges the financial support by BMBF under the Grant Number 03HY201B as part of the public research project "TransHyDE.". FMü gratefully acknowledges financial support by the BMBF under the Grant Number 03SF0693A of the collaborative research project "Energie-Innovationszentrum". The authors would like to thank Friedrich Mendler for his constructive feedback and valuable suggestions, which contributed to the improvement of this paper.

# APPENDIX

## Appendix A.     Insights into Import Costs

This appendix provides a detailed analysis of the distribution of international supply costs for ten exporting countries at the European border. Each exporting country has multiple data points for each commodity and year, linked to the combination of renewables, the quality of these renewables, and their distances from the export node (as discussed in Chapter 4.1.1). In contrast to the supply curves in Chapter 5.1, which display the cheapest sources of commodities in ascending order and are limited by European demand to determine shadow prices, Figure 10 presents the range of supply costs for each exporter and commodity up to a cumulative production of 3.000 TWh. This offers a clearer understanding of the cost differences between countries.

Figure 10 highlights that countries like Norway, Morocco, Chile and United States consistently report low supply costs across all commodities. In these countries, supply costs can still vary significantly due to the different qualities of their renewable resources. For instance, the United States shows considerable variation within its renewable energy qualities, while the United Arabic Emirates has a narrower range, indicating strong and consistent renewable potential across most of its energy sources.

The differences in supply costs between countries are largely driven by two main factors: the quality of renewable resources and country-specific investment costs. These factors are exemplified by capacity factors for renewable energy (see Figure 11) and country-specific WACC (see Figure 12). Figure 11 shows that countries like Chile and Norway have some of the highest capacity factors for offshore and onshore wind, making these locations highly efficient for energy production. Conversely, PV systems generally have lower capacity factors but perform well in countries like Egypt, Oman, and the United Arab Emirates. Despite the favourable renewable profiles in Tunisia and Algeria, these countries report higher supply costs compared to Chile and the United Arab Emirates. This discrepancy can be attributed to variations in WACC, as illustrated in Figure 12. This demonstrates that while renewable resource quality is crucial, country-specific financial conditions, such as WACC, play a significant role in determining the overall supply costs of commodities.

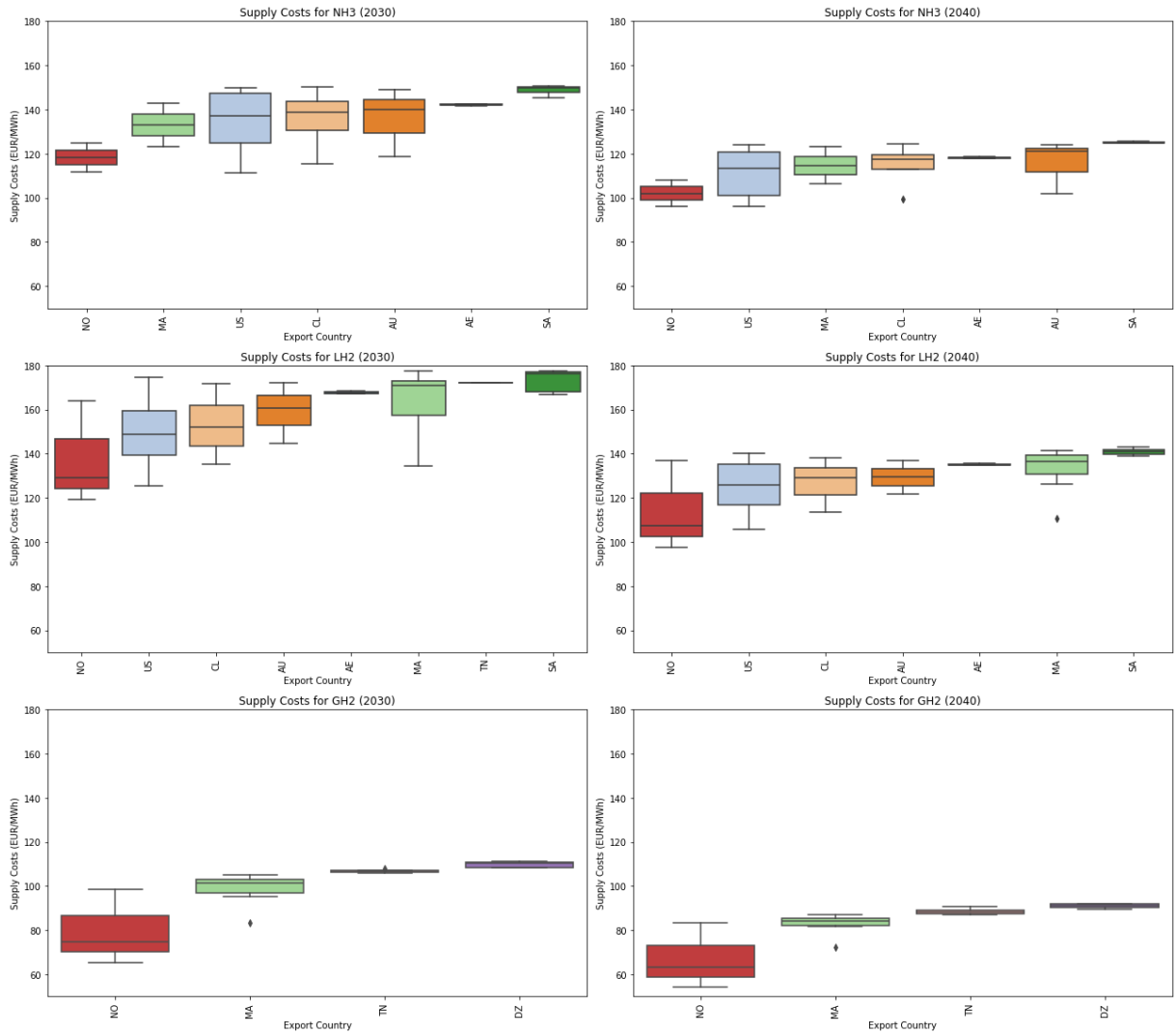

**Figure 10. Distribution of Supply Costs at the European Border (EUR/MWh) for different export countries and commodities (NH3, LH2, GH2) in the years 2030 and 2040. The countries are sorted by the median supply costs, with outliers and variations within each country represented in the boxplots.**

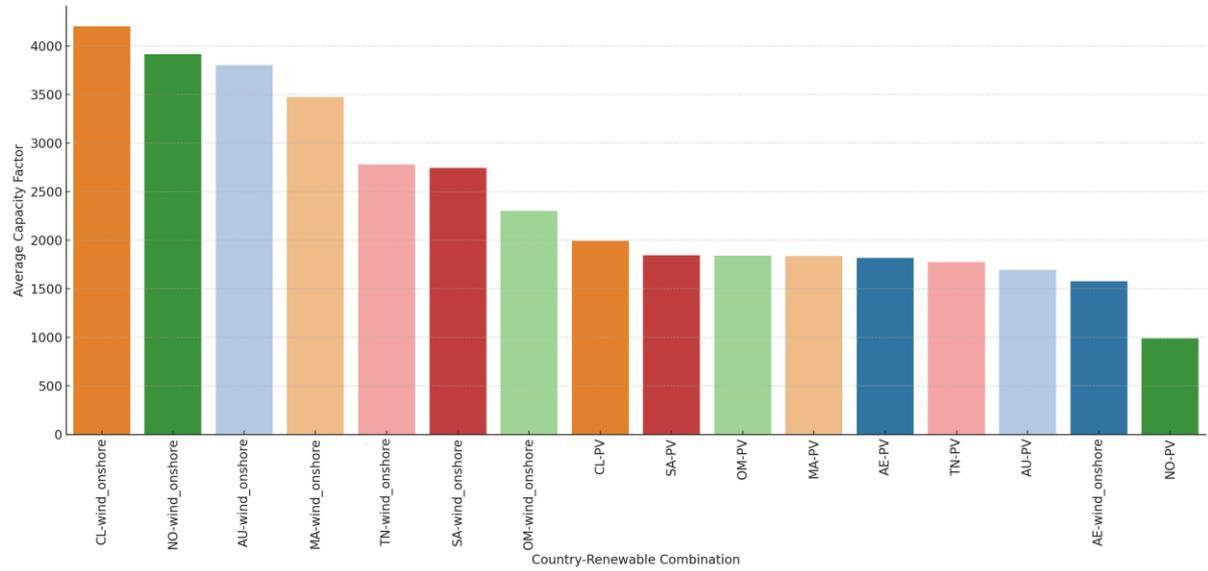

**Figure 11 Average full load hours depending on the exporting country and their renewables.**

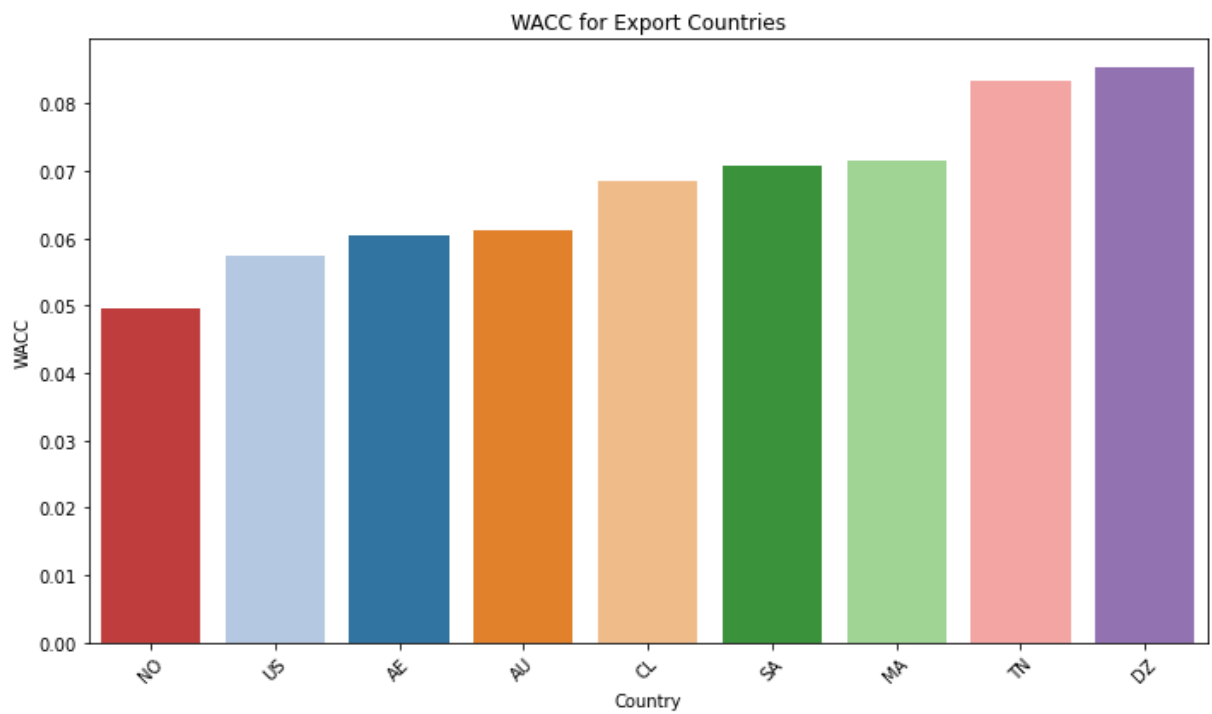

**Figure 12 Country-specific WACC assumed in this study.**

# Appendix B. Analysis of Distribution Modes and Cost Variance for Generic Consumers

This appendix provides a deeper analysis of the Border-to-Consumer model (see Chapter 3.3), expanding upon the supply costs at the consumer site presented in Chapter 5.2. It details the cost composition, distribution mode selection, and cost variance across different demand levels and transport distances.

The following section explains how supply costs vary with demand and transport distance. To provide a clearer focus, this analysis narrows to hydrogen consumers in 2030. The mechanisms, however, remain the same for ammonia consumers and for later years.

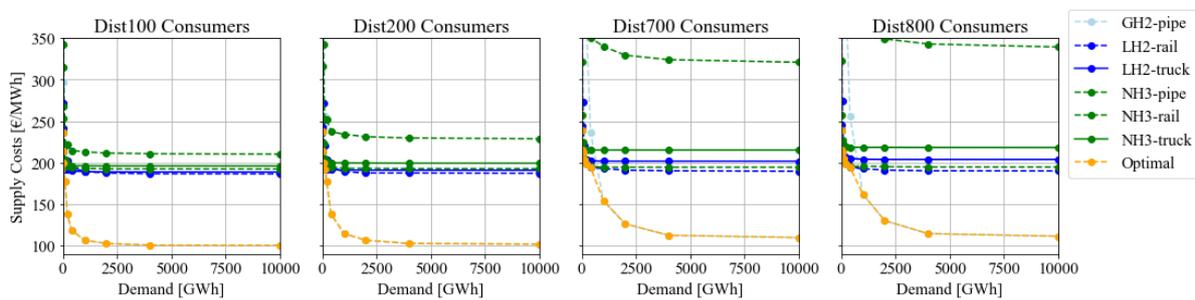

Figure 13. Supply costs as a function of demand for selected distances. X-axis full scale.

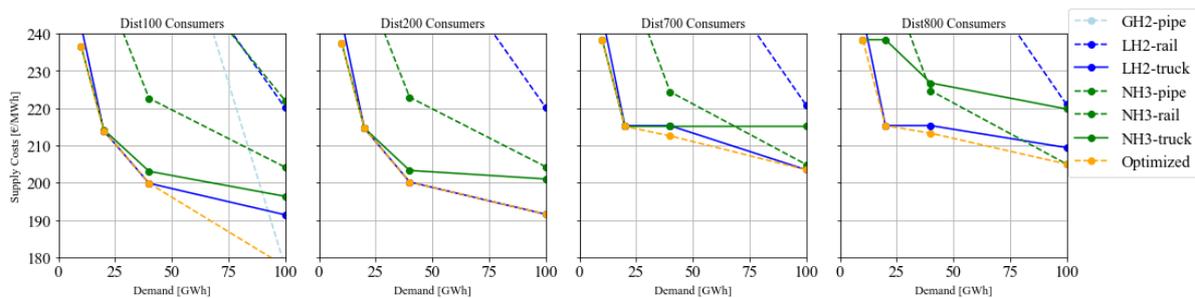

Figure 14. Supply costs as a function of demand for selected distances. X-axis limited to 100 GWh/a.

The analysis reveals clear cost trends as a function of demand, as depicted in Figure 13 and Figure 14, which illustrate supply costs for ten consumers across varying transport distances (100 km, 200 km, 800 km, and 900 km). Supply costs are calculated for each individual distribution mode (represented in blue and green) and for the cost-optimized hybrid mode (represented in orange).

Figure 13 highlights three primary insights. First, supply costs consistently decrease as demand increases due to economies of scale, enhancing efficiency. Second, each distribution mode achieves its minimum cost at distinct demand thresholds, reflecting their inherent efficiencies at varying scales. Finally, the optimal mode selection clearly depends on the specific demand level: truck transport is ideal for small-scale demands below 100 GWh, rail is most cost-effective for medium-scale demands ranging from 100 to 400 GWh, and pipeline transport becomes optimal for large-scale demands exceeding 1,000 GWh.

Figure 14 emphasizes how the model consistently identifies the lowest-cost solution (orange curve) by intelligently combining multiple distribution modes. For example, at distances of 700 km and 800 km with a demand of 40 GWh/a, hybrid distribution modes are employed to minimize total supply costs effectively.

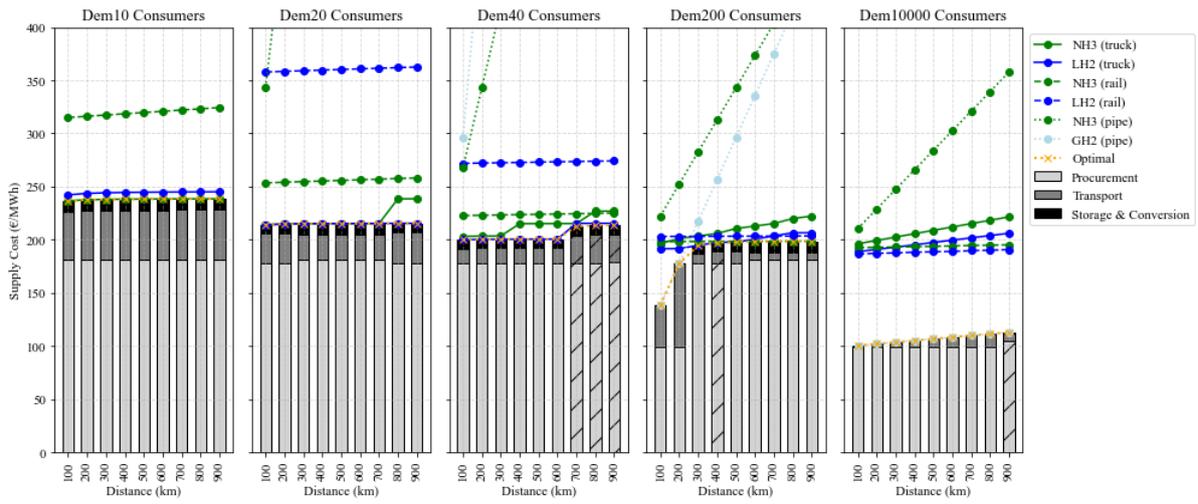

**Figure 15. Supply costs as a function distance for selected demands.**

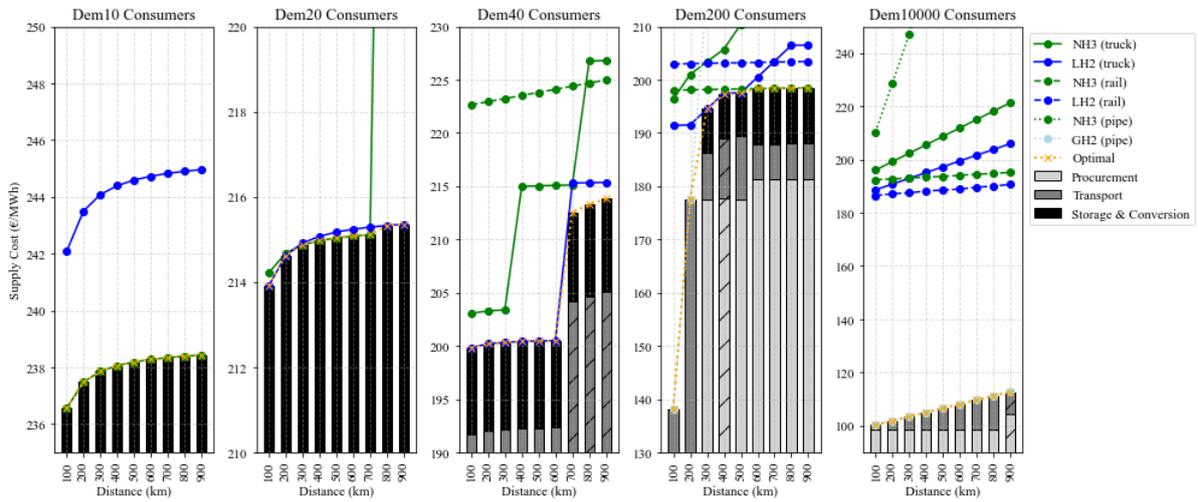

**Figure 16. Supply costs as a function distance for selected demands. Y-axis limited individually.**

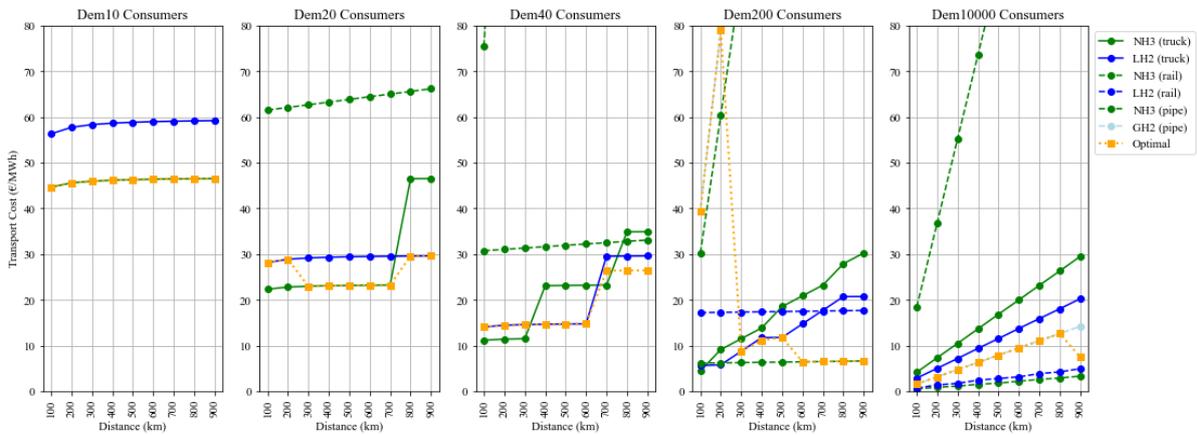

**Figure 17. Transport costs as a function of distance for selected demands.**

The cost trends as a function of distance are detailed in Figure 15 and Figure 16, which present the supply costs for ten consumers at various demand levels (10 GWh/a, 20 GWh/a, 40 GWh/a, 200 GWh/a, and 10,000 GWh/a). Each figure displays three distinct types of curves: blue and green curves represent supply costs when relying on single transport modes (trucks, rail, or pipelines), while the orange curve shows the optimized transport solution, potentially combining multiple transport modes for cost minimization. Notably, abrupt increases or "jumps" in the curves appear whenever an additional transport unit (such as an extra truck or train) is required. Dashed bars further emphasize scenarios where hybrid approaches (integrating two or more distribution modes) are utilized.

The analysis underscores that supply costs rise with increasing distance, significantly impacting the selection of distribution modes. Even with identical demand levels, optimal transport solutions vary according to distance, as demonstrated by distinct transitions highlighted within the figures. For instance, at a demand level of 200 GWh/a, gaseous hydrogen pipelines offer the lowest costs for distances up to 200 km, leveraging their low procurement prices. Between 200 and 400 km, liquid hydrogen trucks become the optimal choice due to their advantageous procurement costs compared to ammonia for hydrogen consumers. Beyond 500 km, ammonia rail transport emerges as the most economical solution, effectively managing long-distance distribution.

Figure 17 specifically isolates transport costs from the same consumers illustrated in Figure 15 and Figure 16, providing two critical insights. Firstly, transport costs alone do not solely dictate the most economical solution, as demonstrated by cases where the lowest transport cost does not correlate with the minimum total cost (e.g., 20 GWh/a demand scenario). Secondly, employing hybrid distribution modes significantly reduces costs, especially notable in scenarios involving 40 GWh/a and 10,000 GWh/a demands.